\documentclass[%
 aps,
 pre,
 amsmath,amssymb,
 reprint,%
longbibliography
]{revtex4-1}

\usepackage{graphicx}
\usepackage{dcolumn}
\usepackage{bm}

\usepackage[utf8]{inputenc}
\usepackage[T1]{fontenc}
\usepackage{mathptmx}
\usepackage{etoolbox}
\usepackage{hyperref}

\makeatletter
\def\@email#1#2{%
 \endgroup
 \patchcmd{\titleblock@produce}
  {\frontmatter@RRAPformat}
  {\frontmatter@RRAPformat{\produce@RRAP{*#1\href{mailto:#2}{#2}}}\frontmatter@RRAPformat}
  {}{}
}%
\makeatother
\begin{document}


\title{Electrical conductivity of crack-template based transparent conductive films: A~computational point of view}
\author{Yuri~Yu.~Tarasevich}
\affiliation{Laboratory of Mathematical Modeling, Astrakhan State University, Astrakhan, 414056, Russia}
\email[Corresponding author: ]{tarasevich@asu.edu.ru}

\author{Andrei~V.~Eserkepov}
\affiliation{Laboratory of Mathematical Modeling, Astrakhan State University, Astrakhan, 414056, Russia}
\email{dantealigjery49@gmail.com}

\author{Irina~V.~Vodolazskaya}
\email{vodolazskaya\_agu@mail.ru}

\affiliation{Laboratory of Mathematical Modeling, Astrakhan State University, Astrakhan, 414056, Russia}

\date{\today}

\begin{abstract}
Crack-template-based transparent conductive films (TCFs) are promising kinds of junction-free, metallic network electrodes that can be used, e.g., for transparent electromagnetic interference (EMI) shielding. Using image processing of published photos of TCFs, we have analyzed the topological and geometrical properties of such crack templates. Additionally, we analyzed the topological and geometrical properties of some computer-generated networks. We computed the electrical conductance of such networks against the number density of their cracks. Comparison of these computations with predictions of the two analytical approaches revealed the proportionality of the electrical conductance to the square root of the number density of the cracks was found, this being consistent with the theoretical predictions.
\end{abstract}

\maketitle

\section{Introduction}\label{sec:intro}

When transparent conductive films (TCFs) are used for electromagnetic interference (EMI) shielding in the optical imaging domain, uniform illumination is crucial for ensuring the imaging quality while the light beam penetrates the metal mesh. The stray light energy from high-order diffractions by the random mesh is significantly less than that from regularly structured meshes (square, honeycomb), which indicates the good optical performance of such random meshes~\cite{Liu2016,Han2016,Shen2018,Jiang2019,Song2023,Li2023}. In contrast to meshes with periodically aligned metal lines, random metal networks produce neither moir\'{e} nor starburst patterns; this property is crucial for their application in displays~\cite{Suh2016,Suh2016a,Jung2019,Walia2019,Melnychenko2022}.

To characterize nanowire-based and templated transparent conductive films, the metal filling factor, $f_\text{F}$, i.e., ratio of the metal-covered area to the total area of the film, is used. The metal fill factor defined in this way and the transmittance, $T$, are connected as follows
\begin{equation}\label{eq:TvsF}
  T = 1 - f_\text{F}
\end{equation}
(see, e.g., \citet{Lee2019}). However, for real-world systems, this relationship is only approximately valid (see Table~\ref{tab:experiments}).
\begin{table*}
\caption{\label{tab:experiments}A summary of experimental data on crack-template-based transparent conductive films. }
\begin{ruledtabular}
\begin{tabular}{ccccccccc}
Source & Metal & Substrate & $R_\Box, \Omega/\Box$ & $T, \%$ & $f_\text{F}$, \% & $w,~\mu$m & $t$, nm & pitch, $\mu$m\\
\hline
\citet{Rao2014a}     & Au & quartz & 5.4   & 87    &  20   & 2 & 60  &  20 \\
\citet{Rao2014a}     & Au & quartz & 3.1   & 87    &  20   & 2 & 220 &  20 \\
\citet{Han2014}      & Ag & glass & 4.2  &  82  & 8.4 & 2 &  60 & 45 \\
\citet{Han2014}      & Ag & PET\footnote{Polyethylene terephthalate} & 10   &  88  & 8.4 & 1 &  60 & 65 \\
\citet{Han2016}      & Ag & quartz &  & 91 &  & 0.5--2 & 200  & 30--80 \\
\citet{Rao2014}      & Ag & glass &  10  &  86   &  20   &   2 & 55  & 20--60  \\
\citet{Pei2015}      & Ag & glass & 6.8   & 86 &  14  & 2   & 50  & 20--80 \\
\citet{Peng2016}     & Ag & PET & 3.7--39.6 & 88.0--99.4  &  0.21--10.25   & 0.5--5  &  10--150 & 40--470 \\
\citet{Voronin2016}  & Ag & PET & 21.4  & 91.6  &       & $14 \pm 0.5$  & 70 &  $78.7 \pm 25.2$ \\
\citet{Voronin2016}  & Ag & PET & 12.3  & 90.1  &       & $14 \pm 0.5$  & 140 &  $78.7 \pm 25.2$ \\
\citet{Voronin2016}  & Ag & PET & 8.1  & 87.4  &       & $14 \pm 0.5$  & 210 &  $78.7 \pm 25.2$ \\
\citet{Voronin2019}  & Ag & PET & 1.3  & 77.3 & $<20$ & 2.4--18.8 & 300 & 40.1--97.5 \\
\citet{Voronin2019}  & Ag & PET & 4.1  & 85.7 & $<20$ & 2.4--18.8 & 300 & 40.1--97.5 \\
\citet{Voronin2021a} & Ag & PET & 11.2 & 90.2 & 10.2 & $3.3 \pm 0.8$ & 200 & $63 \pm 22$ \\
\citet{Voronin2021a} & Ag & PET & 6.8   & 83.6  & 15.5  &  $5.4 \pm 1.4$   & 200 & $67 \pm 25$ \\
\citet{Voronin2021}  & Ag & PET & 11 & 89.8  &  9.8   & 4.7  & 109  & 50 \\
\citet{Voronin2023}  & Ag & PET & 1.59  & 89.1  & 2.9--17  &   1.2--6.2      & 600--2500 & 59.76--94.23\\
\citet{Kang2022}     & Ag & PET & 1.01--5.7   & 88--91.8  & & 0.85--1.53  & 350--750 & 15.0--22.96 \\
\citet{Cheuk2016}    & Ag & glass &  7.8--32.0  &  82 & 9.04--36.8 & 6.6--21.6  & 100 & 55--137\footnote{Computed as square root of the cell area.} \\
\citet{Voronin2022}  & Cu & PET, glass & 2.43  & 91.2  & 7.85  & $7.05 \pm 1.77$ & $443.14 \pm 19.75$ & $177.5 \pm 72.9$ \\
\citet{Voronin2022}  & Cu & PET, glass & 0.53  & 73.8  &  19.5 &  $7.87 \pm 1.09 $ & $723.03 \pm 50.15$ & $76.5 \pm 35.5$ \\
\citet{Walia2019}    & Cu & PET & 0.79  & 90  & 12.5  &          & 300 &  \\
\citet{Walia2020}    & Cu & PET & 0.83--1000  & 77.6--95  & 7.5--25  &   4.5--10       & 20--1000 & 15--35 \\
\citet{Liu2022}  & Cu & PET & 3.4--13.4  & 76.5--93  &   & 1.5  & 120  &  \\
\citet{Mondal2020} & Al & glass & 6 & 91.5  & 9 & 10 & 400 & 100 \\
\citet{Govind2022} & Al & PET & 6.42--55 & 85--95 & & 3--20 & 140--200 &  \\
\end{tabular}
\end{ruledtabular}
\end{table*}

\begin{figure}[!htb]
  \centering
  \includegraphics[width=0.5\columnwidth]{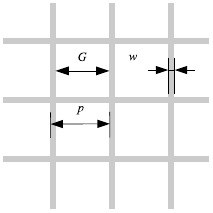}
  \caption{Main sizes used in definitions of the filling factor and of transparency. Here, $p$ is the pitch, $w$ is the line width, $G$ is the grid spacing.}\label{fig:sizes}
\end{figure}

\citet{Ghosh2010,Jiang2019} used the relation
\begin{equation}\label{eq:TvsFJiang2019}
  T = (1 - f_\text{F})^2,
\end{equation}
which assumes an unusual definition of the filling factor, viz., for a regular square grid,
\begin{equation}\label{eq:FFreggrid}
f_\text{F} = \frac{w}{G + w},
\end{equation}
where $w$ is the line width, while $G$ is the grid spacing defined as the shortest distance between two parallel conductive lines (not their pitch!)~\cite{Ghosh2010}.
Hence,
\begin{equation}\label{eq:Treggrid}
T = \left(\frac{G}{G + w}\right)^2,
\end{equation}
This relation has also been used for regular grids of other structures and for irregular meshes~\cite{Jiang2019}.
For a regular square metallic grid, \citet{Schneider2016} estimated the transmittance as follows
\begin{equation}\label{eq:TransmLee}
T = 1 - f_\text{F} = \left(\frac{p-w}{p}\right)^2,
\end{equation}
where $p$ is the pitch. Since $p = G + w$ (see Fig.~\ref{fig:sizes}), the transparencies \eqref{eq:Treggrid} and \eqref{eq:TransmLee} are  identical.

For a regular, square grid, the sheet resistance depends on the fill factor as follows
\begin{equation}\label{eq:RsGrid}
  R_\Box = \xi \frac{\rho}{t f_\text{F}},
\end{equation}
where $\rho$ is the electrical resistivity of the material, $t$ is the thickness of the lines, and $\xi$ is the correction factor~\cite{Ghosh2010,Schneider2016}.
Accounting for~\eqref{eq:TransmLee} and~\eqref{eq:RsGrid},
\begin{equation}\label{eq:TransmvsRsheetLee}
T = 1 - \xi \frac{\rho}{tR_\Box }.
\end{equation}
By contrast, \citet{Muzzillo2020b}, assuming the grids had an idealized square shape, used a quadratic dependence of the transparency on the inverse sheet resistance (i.e., the sheet conductance)
\begin{equation}\label{eq:TvsRsheetsquaremesh}
T = \left(1 - \frac{B \rho}{ t R_\Box}\right)^2,
\end{equation}
where $B$ is the fitting constant.

For irregular metal meshes based on cracked templates, \citet{Voronin2022} defined the fill factor as follows
\begin{equation}\label{eq:FFVoronin}
f_\text{F} = 1 - \left(\frac{g - w}{g}\right)^2,
\end{equation}
where $g$ is the average cell size, while $w$ is the average crack width.
Hence, the transmittance is equal to
\begin{equation}\label{eq:TVoronin}
T = \left(\frac{g - w}{g}\right)^2.
\end{equation}

Using geometrical considerations, \citet{Kumar2016} evaluated the sheet resistances of random metallic networks as
\begin{equation}\label{eq:Kumar}
  R_\Box = \frac{\pi\rho}{2 w t \sqrt{n_\text{E}}},
\end{equation}
where $\rho$ is the resistivity of the material, $w$ and $t$ are the width and the thickness of the rectangular wire, respectively, and $n_\text{E}$ is the number of wire segments per unit area. According to \citet{Kumar2016}, in a sample $L_x \times L_y$, the number of cracks intersecting an equipotential line is
\begin{subequations}\label{eq:intersecKumar}
\begin{equation}
  N_x = \sqrt{n_\text{E}} L_y,
\end{equation}
\begin{equation}
  N_y = \sqrt{n_\text{E}} L_x,
\end{equation}
\end{subequations}
when the potential difference, $U_0$, is applied along the $x$- or $y$-axis, respectively.
The current distribution has also been calculated in conducting crack-template-based metallic networks~\cite{Kumar2017a}.

A similar approach has been applied to anisotropic systems~\cite{Tarasevich2019}. In this case, the sheet resistance can be written as
\begin{equation}\label{eq:Ranisotropic}
  R_\Box =\frac{2 \rho}{ w t n_\text{E} \langle l \rangle (1 \pm s)},
\end{equation}
where $l$ is the length of the crack segment, hereinafter $\langle \cdot \rangle$ denotes an average value,
\begin{equation}\label{eq:s}
s = 2 \langle \cos^2 \theta \rangle - 1
\end{equation}
is the orientational order parameter~\cite{Frenkel1985}, and $\theta$ is the angle between a wire and the $x$-axis. For isotropic systems ($s=0$), equations \eqref{eq:Kumar} and \eqref{eq:Ranisotropic} differ since they are based on different assumptions regarding the number of wires intersecting a line. According to \citet{Tarasevich2019}, the number of cracks intersecting an equipotential line is
\begin{subequations}\label{eq:intersec}
\begin{equation}
  N_x  = \frac{2 \langle l \rangle n_\text{E} L_y}{\pi },
  \end{equation}
\begin{equation}
N_y = \frac{2 \langle l \rangle n_\text{E} L_x}{\pi }.
\end{equation}
\end{subequations}
The average length of the crack segments is expected to be dependent on the number density of the cracks as
\begin{equation}\label{eq:meanlvsnE}
  \langle l \rangle = \beta n_\text{E}^{-1/2}.
\end{equation}
This relation can be easily checked (not proved!) using, e.g., a regular square mesh. The factor $\beta$ depends on the shape factor (circularity) of the cells
\begin{equation}\label{eq:shapefactor}
  C = \frac{4\pi A}{P^2},
\end{equation}
where $A$ is the cell area and $P$ is the cell perimeter.

The crack density is defined as the total crack length, $L_c$, divided by the area of the reference surface, $A$~\cite{Roux2013},
\begin{equation}\label{eq:crackdens}
  \rho_\text{c} = \frac{L_c}{A}.
\end{equation}

In general terms, omitting technical details, the manufacture of crack-template-based TCFs consists of the following steps. A treated substrate (glass, quartz, PET, etc.) is covered (using, e.g., the Meyer rod method or by spin coating) width a thin film of a polymer or colloid (e.g., egg white). This thin film cracks due to desiccation, producing a crack template. A metal (Au, Ag, Cu, Al, etc.) is sputtered onto the template. Then, the template may be removed (e.g., dissolved). The metal mesh on the substrate is used as a seed for the galvanic deposition of the same or another metal. Although the actual technology may differ significantly, this description offers a basic idea of the manufacture of crack-template-based TCFs. Detailed descriptions of the manufacturing of each particular sample can be found in the appropriate references presented in Table~\ref{tab:experiments}.

Basic information regarding the formation and modelling of desiccation crack patterns can be found in \citet{Goehring2015}. \citet{Zeng2020} proposed a coupled electrothermal model to describe the physical properties of crack-template-based transparent conductive films.  The geometry of the random metallic network was generated by applying a Voronoi diagram. In particular, the current density and heating power have been computed using COMSOL. \citet{Kim2022} proposed a geometric modeling approach for crack-template-based networks. The authors mimicked real-world crack patterns using a Voronoi tessellation. Within Model~I, wires of \emph{varying} width were  assigned to the edges of the polygonal cells of the Voronoi tessellation. The resulting geometry in Model~I is hyperuniform. Within Model~II, wires of \emph{equal} width were assigned to the edge of each Voronoi polygon. The authors  computed the optical transmittance and the sheet resistance of these networks. In fact, both these models use an assumption that the width of the cracks was distributed independently and they ignored the hierarchy of the cracks. However, the widths of the any adjacent segments of the same crack are not independent; besides, primary cracks are wider than secondary ones. Nevertheless, when the mean electrical properties are of interest, the fine structure of any particular crack pattern might effectively be negligible. This issue needs to be additionally studied, at least. \citet{Esteki2023}  mimicked seamless metallic nanowire networks using a Voronoi tessellation. A computational study of the thermo-electro-optical properties of these networks and their geometrical features was performed using in-house computational implementations and a coupled electrothermal model built in COMSOL Multiphysics software.

Although an extensive comparison of real-world crack patterns and mosaics (Gilbert and Voronoi tessellations) has recently been performed~\cite{Roy2022}, we have applied a simpler analysis to find some common features of the crack patterns used in TCE production.

The goal of the present work is an investigation of the topological, geometrical and electrical properties of TCFs obtained using crack templates, as well as of artificial computer-generated networks that are intended to mimic the properties of interest of these real-world TCFs.

The rest of the paper is constructed as follows. Section~\ref{sec:methods} describes technical details of the image processing and simulation, together with the analytical approach, but also presents some preliminary results. Section~\ref{sec:results} presents our main findings. Section~\ref{sec:concl} summarizes the main results. Some mathematical details are presented in Appendix~\ref{sec:variablewidth}, Appendix~\ref{sec:cond}, and Appendix~\ref{sec:relationship}.

\section{Methods}\label{sec:methods}

\subsection{Sampling}

We studied 10 real crack-template-based networks and two kinds of computer-generated networks. We refer to the real-world networks as sample~1 and sample~2, and these correspond to the samples\footnote{The unpublished photos were kindly provided by A.S.~Voronin who is one the authors of that article\cite{Voronin2021}} described in Ref.~\onlinecite{Voronin2021}, while sample~3 corresponds to~\citet[cropped Fig. 2(a)]{Voronin2016}; sample~4, which corresponds to \citet[cropped Fig. 2(a)]{Xian2017}, while sample~5, sample~6, and sample~7 correspond to \citet[Fig. 2, cropped subfigures (a), (b), and (c), correspondingly]{Gao2018}; sample~8, sample~9, and sample~10 correspond to \citet[Fig. 2, cropped subfigures (a), (b), and (c), respectively]{Pei2015}.

Table~\ref{tab:experiments} shows that there is a significant spread in materials and parameters used; in the available literature, the values of sheet resistances are not always related to specific values of geometric parameters, while in cases where photographs of the samples are given, their specific electrical and geometric characteristics are often missing. For these reasons, we are unable directly to compare the results of our computations with available experimental data. Instead, for analysis, we have used the reduced electrical conductance.

\begin{figure*}
  \centering
  \includegraphics[width=\textwidth]{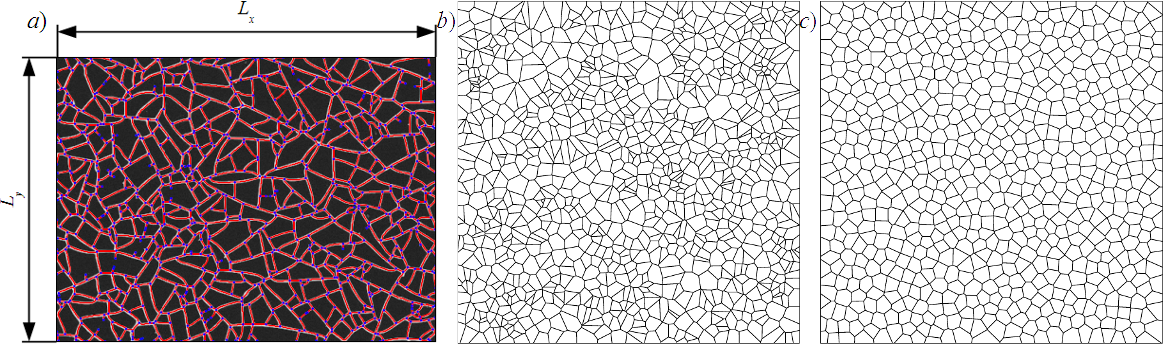}\\
  \caption{Examples of the crack patterns. a)~Sample~1, a real-world crack pattern (courtesy of A.S.~Voronin) along with a corresponding network obtained using StructuralGT~\cite{Vecchio2021}. b)~A computer-generated network produced using Voronoi tessellation based on randomly distributed points. c)~A computer-generated network produced using Voronoi tessellation based on random hyperuniformly distributed points.\label{fig:samples}}
\end{figure*}

We refer to the computer-generated networks as Voronoi and VoronoiHU. Figure~\ref{fig:samples}(b) and fig.~\ref{fig:samples}(c) demonstrate examples of such computer-generated networks.  These networks were produced by applying the Voronoi tessellation in MATLAB. In the first case, the points were randomly distributed within a square domain. In the second case, discs were deposited in the square domain using  random sequential adsorption. The centers of these discs were used as random hyperuniformly distributed points in order to perform a Voronoi tessellation. Note, that the VoronoiHU networks resemble printed random meshes\cite{Liu2016,Shen2018,Li2023} rather than crack-template-based networks.

\subsection{Image processing}\label{subsec:image}
To process the images of crack patterns, we used StructuralGT, a Python package for automated graph theory analysis of structural network images~\cite{Vecchio2021}. StructuralGT was modified to match our particular requirements. Figure~\ref{fig:samples}(a) demonstrates an example of a crack pattern along with its corresponding network. This network, which mimics the crack pattern, was generated using StructuralGT~\cite{Vecchio2021}. The sample size is $1.65 \times 1.2365$~mm. The network of this particular crack pattern contains 899 nodes and 1249 edges. The average degree is $\deg V \approx 2.78$. Nodes with $\deg V =1$ correspond to dead ends as well as to the intersections of edges with the domain boundaries, while ones with $\deg V =2$ correspond to bent cracks. The overall length of edges, which are incident on nodes with $\deg V =1$, is about 7.4\% of the total length of edges. Analysis of the crack widths was omitted due to the modest resolution of the images, which would have lead to insufficient accuracy.

\subsection{Computational details}\label{subsec:compute}
The networks under consideration (both the real-world samples and computer-generated networks) were treated as random resistor networks (RRNs), where each network edge corresponded to a resistor, while each node corresponded to a junction between resistors. The resistance of an $i$-th resistor can be written as
\begin{equation}\label{eq:resistor}
  R_i = \rho \frac{l_i}{A_i},
\end{equation}
where $l_i$ is the crack length, while $A_i = w_i t_i$ is the area of the cross-section of the metal that  fills the crack. We supposed that all such metal wires have the same width and thickness. The impact of variable width of a conductor on the electrical resistance is considered in Appendix~\ref{sec:variablewidth}.

\begin{figure}[!htb]
  \centering
  \includegraphics[width=\columnwidth]{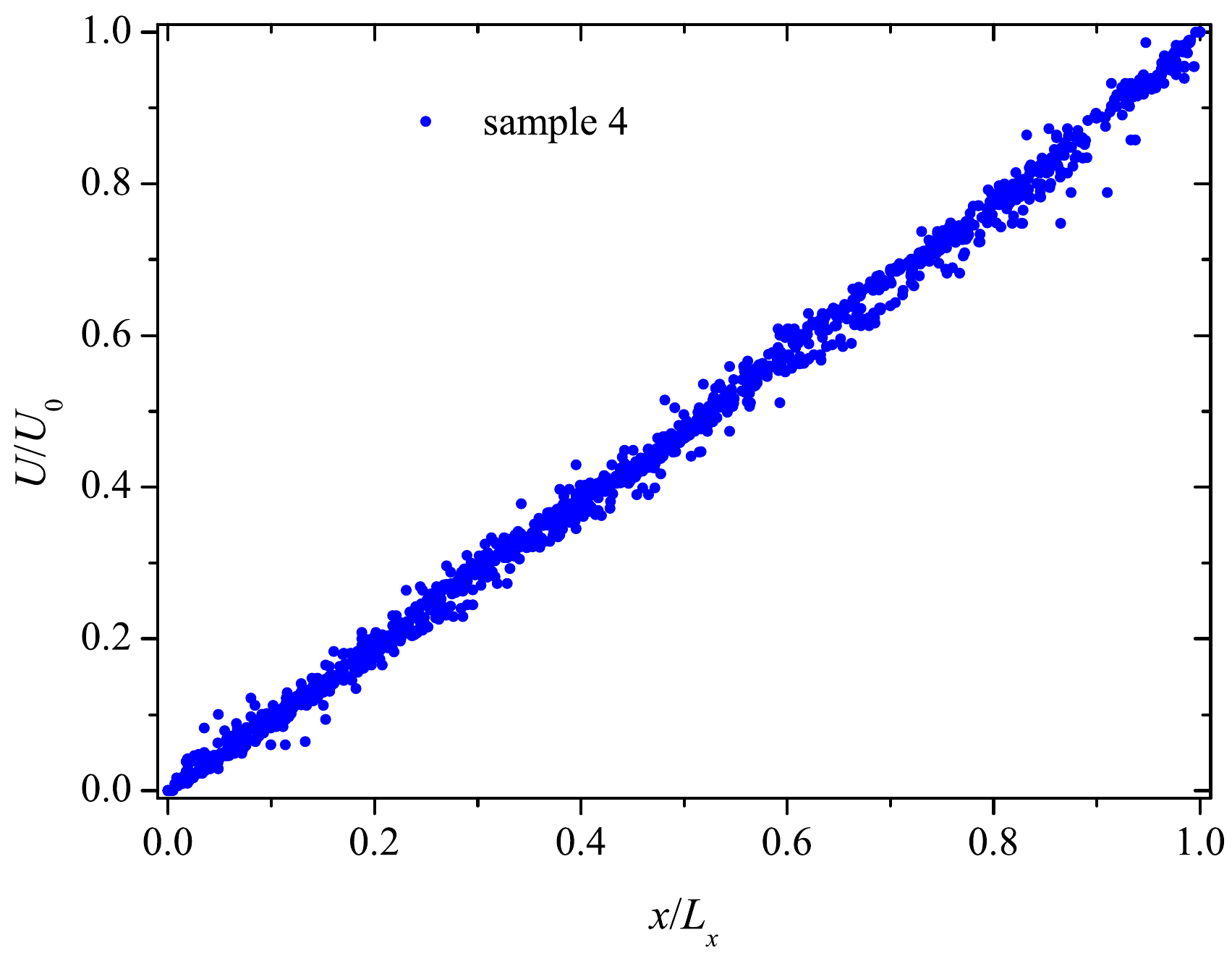}
  \caption{Example of a potential distribution along a sample.}\label{fig:Vvsxsample4}
\end{figure}

We attached a pair of superconducting buses to the two opposite boundaries of the systems in such a way, that the potential difference, $U_0$, was applied either along axis $x$ or along axis~$y$. Applying Ohm's law to each resistor and Kirchhoff's point rule to each junction, a system of linear equations (SLEs) was obtained. The matrix of this SLE is sparse. Such an SLE can be solved numerically to find the potentials and currents in the RRN under consideration. We used the EIGEN library\cite{Guennebaud2010} to solve the SLEs. In particular, our computations evidenced that the potential drop along the samples is close to linear (Fig.~\ref{fig:Vvsxsample4}). When the current in each resistor is known, the total electric current can be calculated. Since the applied potential difference, $U_0$, is known, the total resistance and the conductance of the RRN can be found using Ohm's law. We use subscripts $x$ and $y$ to distinguish the resistances and conductances along axis $x$ or along axis~$y$. Then, the sheet resistance and conductance could be found, since
\begin{equation}\label{eq:RvsRsq}
  R_\Box = R_x\frac{L_y}{L_x} = R_y\frac{L_x}{L_y}.
\end{equation}

\subsection{Theoretical description}\label{subsec:theory}
A geometrical consideration has been used to evaluate the electrical conductivity of crack-template-based TCFs~\cite{Kumar2016}. In fact, this consideration is a kind of mean-field approach, since a single wire (crack segment) is considered to be placed in a homogeneous electric field produced by all the other wires. In our study, we used a very similar idea and treated as crack-template-based TCFs as random networks.

Since homogeneous cracking structures without clusters and blind cracks (dead ends) can be produced~\cite{Voronin2016,Voronin2021}, we suppose that
\begin{enumerate}
  \item the template consists of a single cluster, i.e., there is only one connected component in the corresponding network;
  \item there are no dead ends, i.e., $\deg V >1$ for any node  in the corresponding network;
  \item the crack template is isotropic, i.e., all crack orientations are equiprobable.
\end{enumerate}
These assumptions will be validated in Section~\ref{sec:results} using a computer simulation.

The sheet resistance is
\begin{equation}\label{eq:Rsheet}
R_\Box = \frac{2 \rho}{ n_\text{E}  \langle l \rangle w t}.
\end{equation}
A detailed derivation of formula~\eqref{eq:Rsheet} can be found in Section~\ref{sec:cond}.
Although this approach is similar to that described in Ref.~\onlinecite{Kumar2016}, the formula for the sheet resistance~\eqref{eq:Rsheet} differs from~\eqref{eq:Kumar} obtained by~\citet{Kumar2016}. This difference is due to the number of edges intersecting any equipotential, viz., to find this number, we used a rigorous probabilistic derivation, while~\citet{Kumar2016}  utilized an estimate ($L_y \sqrt{n_\text{E}}$). It would be very tempting to simplify formula~\eqref{eq:Rsheet} by finding the exact value of the factor $\beta$ in the relationship between the average length of a crack segment and the concentration of cracks~\eqref{eq:meanlvsnE}. Unfortunately, only estimates are possible (see Appendix~\ref{sec:relationship}). However, Figure~\ref{fig:lmeanvsn} evidenced that~\eqref{eq:meanlvsnE} holds for the networks under consideration.
\begin{figure}[!htb]
  \centering
  \includegraphics[width=\columnwidth]{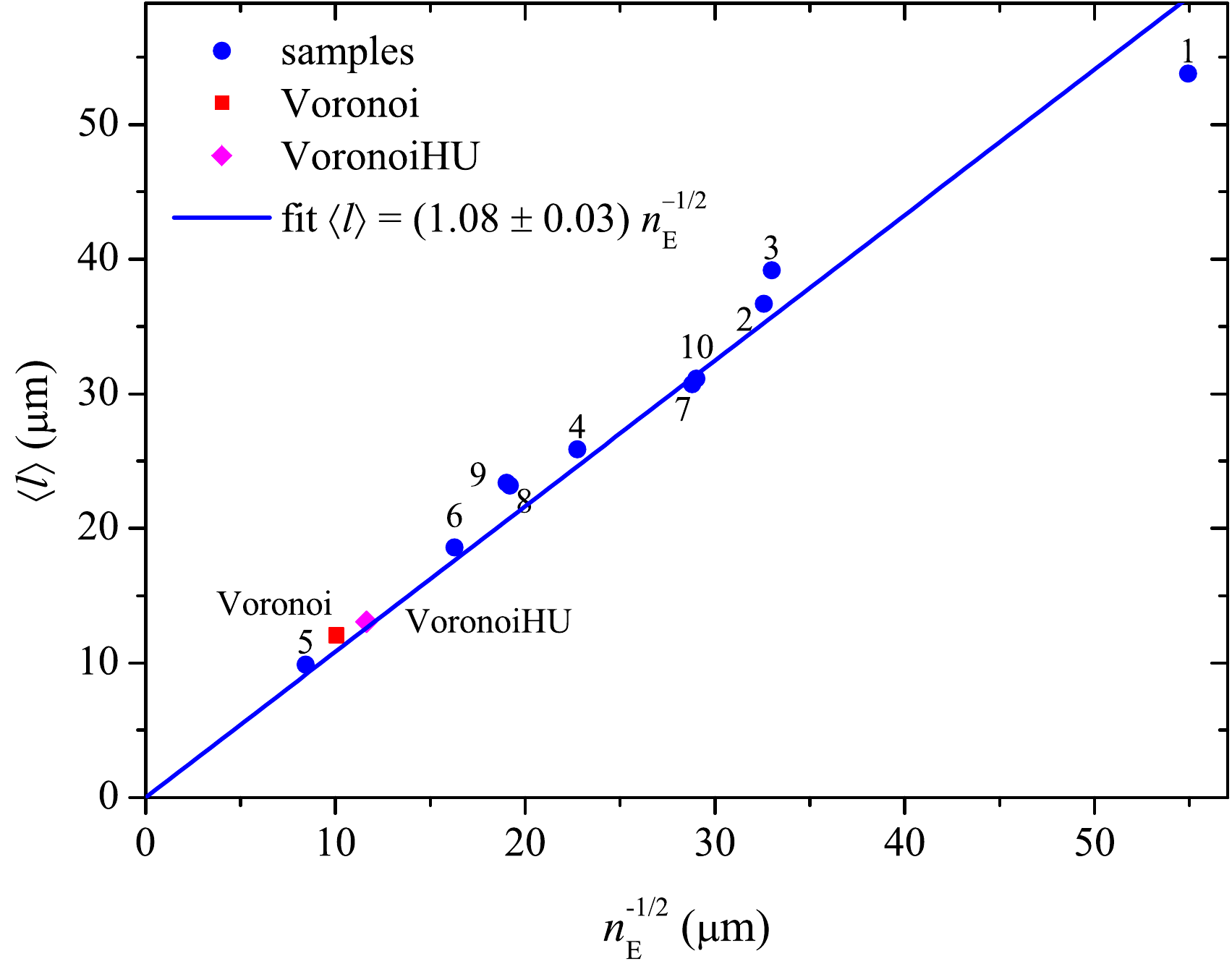}
  \caption{Dependency of the mean length of crack segments on the number density of the cracks.}\label{fig:lmeanvsn}
\end{figure}

Note, the dependence of the sheet conductance on the fill fraction is  linear, viz.,
\begin{equation}\label{eq:GvsF}
  G_\Box = \frac{\sigma t f_\text{F}}{2},
\end{equation}
since ${f_\text{F}} = n_\text{E} w \langle l \rangle$. Accounting for~\eqref{eq:TvsF}, $n_\text{E} w \langle l \rangle = 1 - T$, hence,
\begin{equation}\label{eq:TvsRsheet}
T = 1 - \frac{2 \rho}{ t R_\Box}.
\end{equation}
This relation resembles that obtained for a square mesh~\eqref{eq:TransmvsRsheetLee}.

The total mass of the metal deposited in the cracks is
\begin{equation}\label{eq:m}
m = n_\text{E}  \langle l \rangle w t \rho_m L_x L_y,
\end{equation}
where $\rho_m$ is the metal density.

\section{Results}\label{sec:results}

\begin{figure}[!htb]
  \centering
  \includegraphics[width=\columnwidth]{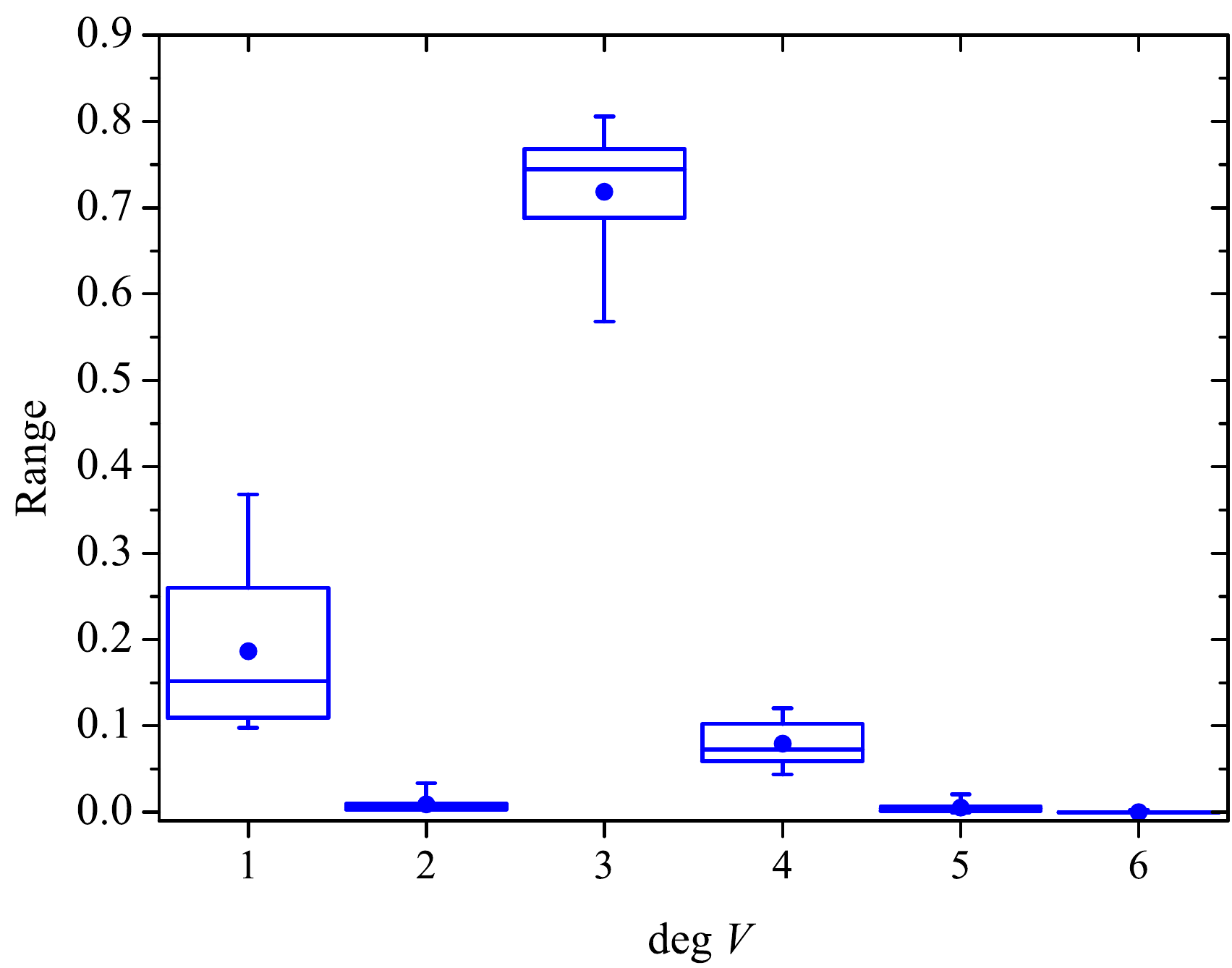}
  \caption{Degree distribution in the 10 real-world samples.\label{fig:degree2p}}
\end{figure}

Figure~\ref{fig:degree2p} demonstrates the degree distribution in these 10 particular samples. In the box and whisker charts, here and below, the mean values are shown using markers, the `box' presents the median, 25 and 75 percentiles, while the `whiskers' indicate the minimal and maximal values. These distributions evidenced that T- and Y-shaped ($\deg V = 3$) crack junctions dominate. $\deg V =1$ corresponds to both the internal nodes (dead ends) and to nodes on the boundaries (the intersections of cracks with image boundaries). A proportion of dead ends is 0.15--0.3, while that of X-shaped ($\deg V = 4$) junctions is about 0.1. Here, we use the same notation to classify junctions as~\citet{Gray1976}. The proportions of nodes corresponding to $\deg V = 2$ and to $\deg V = 5$ are negligible. We suppose that vertices owning both $\deg V = 2$ and $\deg V > 4$ are artefacts rather than a reality, since the modest resolution of the images leads to only moderate accuracy of the image processing. By contrast, in networks produced using Voronoi tessellation, nodes with $\deg V =1$ correspond solely to the intersections of edges with the domain boundaries. All internal nodes match $\deg V = 3$. In general, in such artificial networks, the distribution of node degrees resembles that in real-world crack patterns.

\begin{figure}[!htb]
  \centering
  \includegraphics[width=\columnwidth]{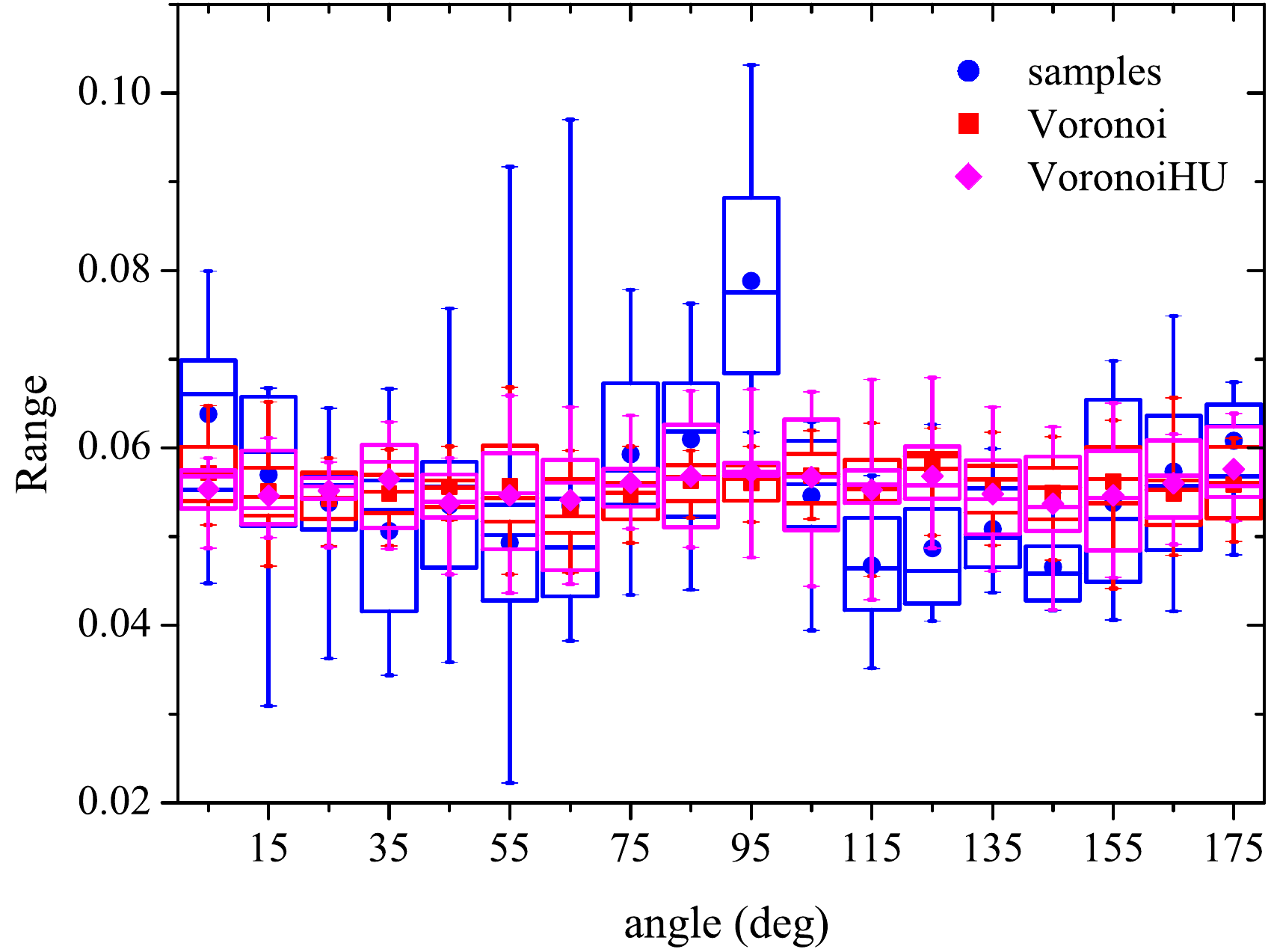}
  \caption{Distribution of crack orientations in the particular real-world samples and in  computer-generated networks produced using Voronoi tessellation.\label{fig:angles}}
\end{figure}

Figure~\ref{fig:angles} demonstrates the distributions of crack orientations in the same real-world samples along with artificial computer-generated networks. These distributions evidenced that the cracks are approximately equiprobably oriented, i.e., the value of the order parameter~\eqref{eq:s} is close to 0. However, the statistical variations are significant. Equiprobable distribution of crack orientations was also found for networks obtained using Voronoi tessellation. The values of the order parameter are presented in Table~\ref{tab:intersect}. However, `brickwork' patterns with two mutually perpendicular sets of parallel cracks can also be produced~\cite{Xie2018}. In that case, the orientations of the cracks have to obey a bimodal distribution.

\begin{figure}[!htb]
  \centering
  \includegraphics[width=\columnwidth]{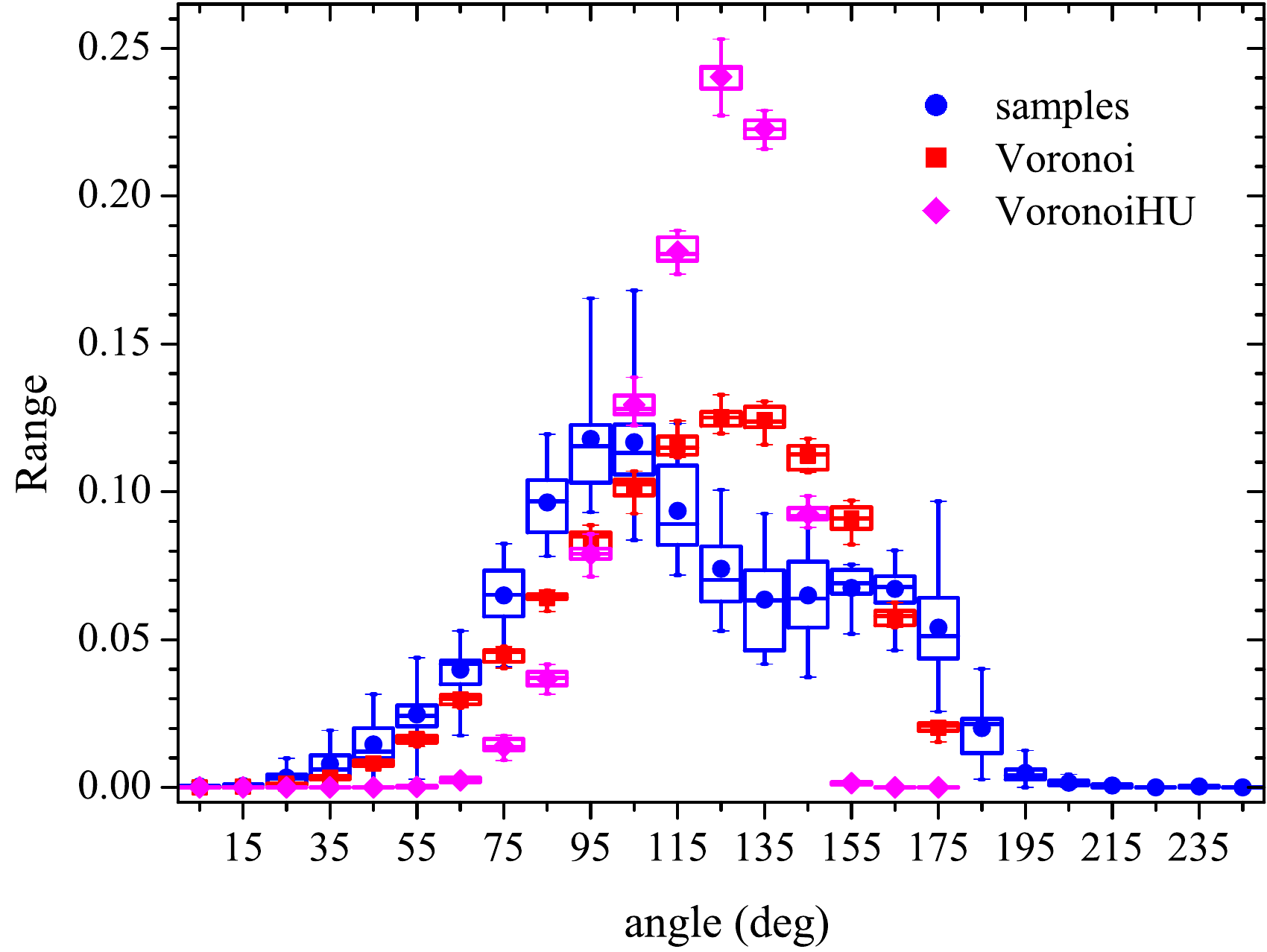}
  \caption{Distribution of angles between adjacent cracks in the real-world samples and in the computer-generated networks produced using Voronoi tessellation.\label{fig:CrackAnglesBox}}
\end{figure}

Figure~\ref{fig:CrackAnglesBox} demonstrates the distributions of angles between adjacent cracks in the same  samples. These distributions evidenced that, in the real-world samples, cracks tend to join at right angles (T-shaped junctions). The maximum of the distribution is located near the angle $90^\circ$ ($\pi/2$~rad) (two right angles in each T-shaped junction); a less pronounces maximum located near $180^\circ$ ($\pi$~rad) corresponds to the straight angles in these T-shaped junctions. Such behavior is quite to be expected~\cite{Bohn2005,Bohn2005a,Kumar2021}. By contrast, networks obtained using Voronoi tessellation demonstrate unimodal distribution with the maximum near $120^\circ$ (2~rad), i.e., here, Y-shaped junctions dominate (Fig.~\ref{fig:CrackAnglesBox}). Note that, in the crack-template-based networks, the cracks are curved, while, in the corresponding networks produced using StructuralGT, the corresponding edges are straight. Consequently, the resulting distribution of angles between adjacent cracks is more diffuse in the vicinity of $90^\circ$ ($\pi/2$~rad) than it should be.

\begin{figure}[!htb]
  \centering
  \includegraphics[width=\columnwidth]{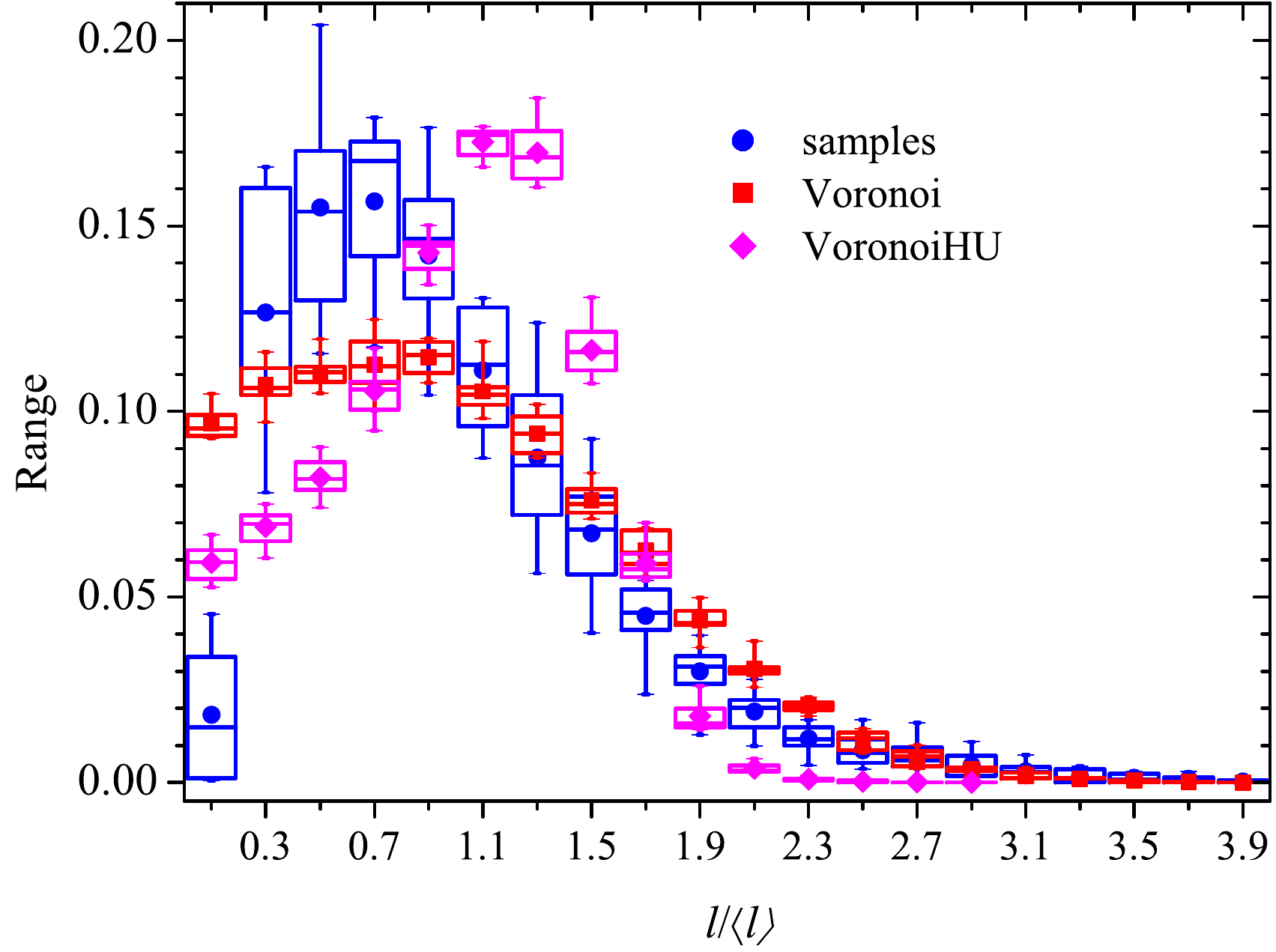}
  \caption{Comparison of distribution of crack lengths in real-world samples and in computer-generated networks produced using Voronoi tessellation.\label{fig:length2p}}
\end{figure}

Figure~\ref{fig:length2p} demonstrates the distributions of normalized crack length in the same samples. Here, $\langle l \rangle$ denotes the mean crack length. Comparison of the distributions evidenced that, in the real-world and in the computer-generated networks, the length distributions are similar and resemble a log-normal distribution.

This preliminary analysis evidenced that Voronoi tessellation produces networks, the topology which is close to the real crack-template-based networks while their geometries differ.

Tables~\ref{tab:intersect} and \ref{tab:intersectVHU} evidenced that the approach by \citet{Kumar2016} significantly (1.5--2 times) overestimates the number of intersections of cracks with an equipotential line both for the real-world networks and for the computer-generated networks. Although the approach by \citet{Tarasevich2019} is better (the overestimation is about 10\% in the case of the real-world samples, while, in the case of the computer-generated networks, the prediction falls within the statistical error), the source of this overestimate needs to be identified.
\begin{table*}[tb]
\caption{\label{tab:intersect} Number of intersections in real-world samples.}
\begin{ruledtabular}
\begin{tabular}{lccccccc}
Smp & $s$ & \multicolumn{2}{c}{Simulation} & \multicolumn{2}{c}{Eq.~\eqref{eq:intersecKumar}} & \multicolumn{2}{c}{Eq.~\eqref{eq:intersec}}\\
 & & $x$ & $y$ & $x$ & $y$ & $x$ & $y$ \\
\hline
 1 & $-0.01  \pm 0.03  $ & $13.08 \pm 0.12$ & $19.35 \pm 0.32$ & 21.4886 & 31.6446 & 14.0815 & 20.7367 \\
 2 & $-0.047 \pm 0.021 $ & $19.35 \pm 0.12$ & $28.16 \pm 0.15$ & 29.9026 & 39.8962 & 21.4941 & 28.6775 \\
 3 & $ 0.007 \pm 0.017 $ & $25.81 \pm 0.18$ & $35.18 \pm 0.20$ & 37.5066 & 49.9112 & 28.2236 & 37.5579 \\
 4 & $ 0.012 \pm 0.017 $ & $35.42 \pm 0.17$ & $35.78 \pm 0.19$ & 50.8029 & 50.9026 & 38.8360 & 38.9122 \\
 5 & $-0.079 \pm 0.023 $ & $16.88 \pm 0.14$ & $28.40 \pm 0.21$ & 25.4009 & 37.0853 & 18.7475 & 27.3714 \\
 6 & $-0.010 \pm 0.022 $ & $17.02 \pm 0.12$ & $29.21 \pm 0.19$ & 25.6971 & 42.2226 & 18.6698 & 30.6761 \\
 7 & $-0.018 \pm 0.030 $ & $13.38 \pm 0.09$ & $18.56 \pm 0.13$ & 20.5013 & 29.5103 & 13.9044 & 20.0145 \\
 8 & $ 0.072 \pm 0.018 $ & $25.49 \pm 0.15$ & $32.90 \pm 0.24$ & 33.3887 & 49.5977 & 25.6134 & 38.0478 \\
 9 & $ 0.046 \pm 0.018 $ & $25.17 \pm 0.16$ & $34.30 \pm 0.24$ & 33.1937 & 51.2447 & 25.9509 & 40.0632 \\
10 & $ 0.006 \pm 0.024 $ & $16.63 \pm 0.14$ & $23.98 \pm 0.26$ & 26.1745 & 38.4344 & 17.8347 & 26.1883 \\
\end{tabular}
\end{ruledtabular}
\end{table*}

\begin{table}[!htb]
\caption{\label{tab:intersectVHU} Number of intersections in computer-generated samples.}
\begin{ruledtabular}
\begin{tabular}{lcccc}
Smp &  \multicolumn{2}{c}{Simulation}  & Eq.~\eqref{eq:intersecKumar} & Eq.~\eqref{eq:intersec}\\
 & $x$ & $y$ &  &  \\
\hline
\multicolumn{5}{c}{Voronoi} \\
\hline
 1 & $39.59 \pm 0.66$ &  $39.47 \pm 0.67$ & 54.2679  & 39.5815 \\
 2 & $39.44 \pm 0.64$ &  $40.63 \pm 0.57$ & 54.3047  & 39.9485 \\
 3 & $40.03 \pm 0.79$ &  $39.75 \pm 0.62$ & 54.3967  & 39.7988 \\
 4 & $38.81 \pm 0.68$ &  $39.69 \pm 0.65$ & 54.3323  & 39.4767 \\
 5 & $38.78 \pm 0.68$ &  $39.16 \pm 0.87$ & 54.4243  & 39.5114 \\
 6 & $39.66 \pm 0.80$ &  $40.13 \pm 0.75$ & 54.4151  & 39.5292 \\
 7 & $40.06 \pm 0.81$ &  $40.34 \pm 0.72$ & 54.3691  & 39.8669 \\
 8 & $39.94 \pm 0.72$ &  $40.03 \pm 0.68$ & 54.3415  & 39.6991 \\
 9 & $39.19 \pm 0.47$ &  $39.84 \pm 0.73$ & 54.3599  & 39.6767 \\
10 & $38.69 \pm 0.78$ &  $40.06 \pm 0.65$ & 54.3783  & 39.6188 \\
\hline
\multicolumn{5}{c}{VoronoiHU} \\
\hline
 1 & $ 31.50 \pm 0.44 $ & $ 31.75 \pm 0.40 $ & 44.5870 & 31.3162 \\
 2 & $ 31.41 \pm 0.45 $ & $ 31.28 \pm 0.38 $ & 44.3959 & 31.2962 \\
 3 & $ 31.00 \pm 0.44 $ & $ 31.75 \pm 0.48 $ & 44.7661 & 31.4797 \\
 4 & $ 31.53 \pm 0.45 $ & $ 31.03 \pm 0.44 $ & 45.0222 & 31.6740 \\
 5 & $ 31.03 \pm 0.37 $ & $ 30.19 \pm 0.46 $ & 44.5197 & 31.3243 \\
 6 & $ 31.19 \pm 0.35 $ & $ 32.03 \pm 0.40 $ & 44.8665 & 31.6408 \\
 7 & $ 31.60 \pm 0.50 $ & $ 32.09 \pm 0.43 $ & 44.9110 & 31.5567 \\
 8 & $ 31.66 \pm 0.41 $ & $ 31.31 \pm 0.49 $ & 44.7661 & 31.5408 \\
 9 & $ 31.34 \pm 0.51 $ & $ 32.06 \pm 0.45 $ & 44.6542 & 31.3690 \\
10 & $ 31.81 \pm 0.39 $ & $ 31.00 \pm 0.47 $ & 44.8553 & 31.5963 \\
\end{tabular}
\end{ruledtabular}
\end{table}

Figure~\ref{fig:lmeanvsn} suggests that, for random networks,  $ \langle l \rangle \approx n_\text{E}^{-1/2}$ since the factor $\beta$ is close to unity. In this case, \eqref{eq:Gsheet} transforms in
\begin{equation}\label{eq:Gsheetsimple}
 G_\Box = \frac{  w t \sigma \sqrt{n_\text{E}}}{2},
\end{equation}
that differs from \eqref{eq:Kumar} by a factor of $\pi/4$.

\begin{figure}[!htb]
  \centering
  \includegraphics[width=\columnwidth]{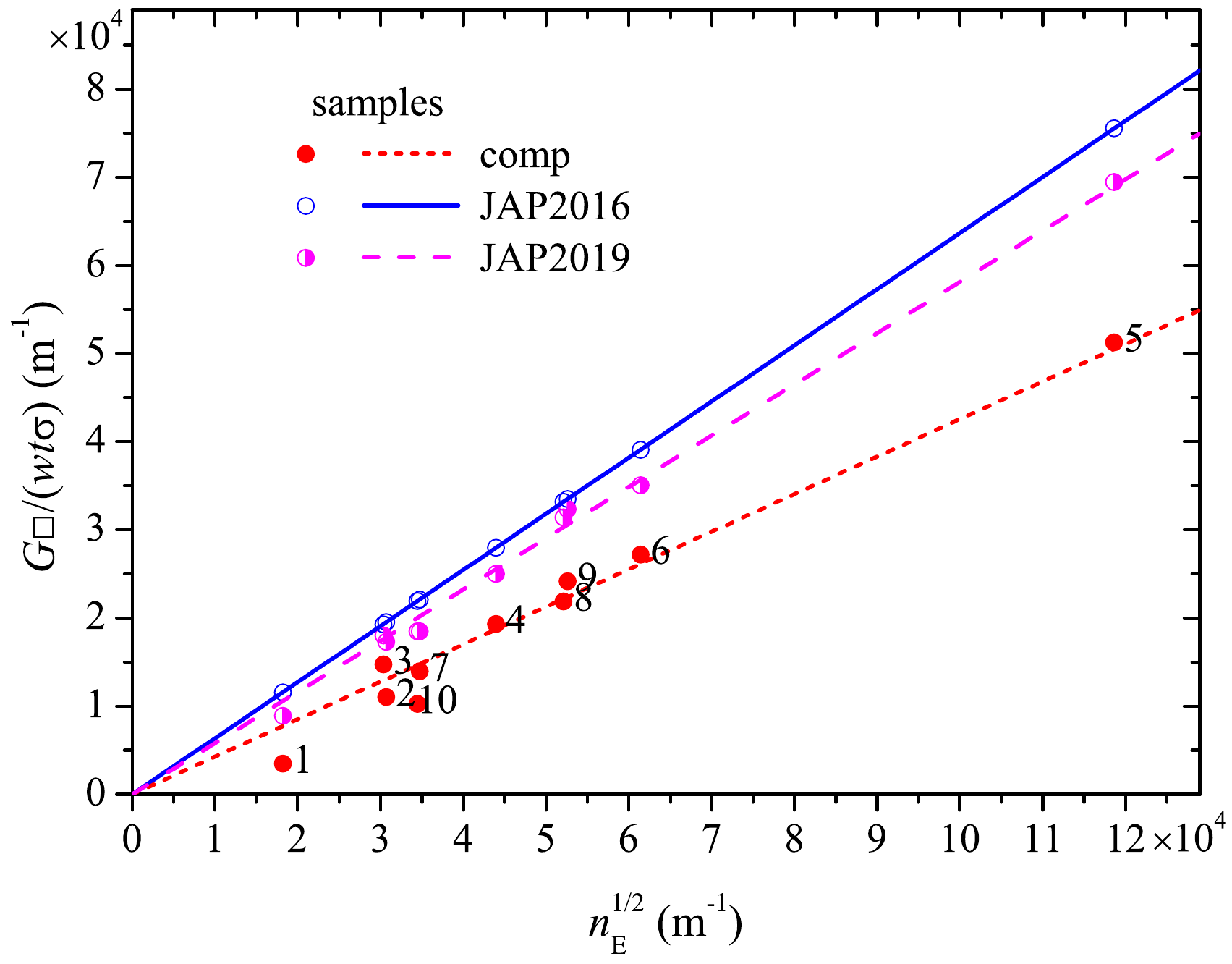}
  \caption{Reduced conductance vs square root of the number density of cracks. JAP2016 corresponds to  prediction~\eqref{eq:GKumar}, JAP2019 corresponds to prediction~\eqref{eq:Gsheet}. Dashed lines correspond to least squares fits.  Numbers near the markers indicate the number of the sample.}\label{fig:Gsqcompar}
\end{figure}

Figure~\ref{fig:Gsqcompar} plots the reduced electrical conductance of the samples against the square root of the number density of the cracks. The numbers near the markers indicate the numbers of the real-world samples. Since there was a difference in sheet resistance of the same sample computed using resistances in two directions, the averaged sheet resistance is presented for each sample. This difference arose due to the limited precision of the image processing caused by the quality of the photos. JAP2016 (markers and solid line) corresponds to prediction
\begin{equation}\label{eq:GKumar}
 \frac{G_\Box}{w t \sigma} = \frac{2}{\pi}\sqrt{n_\text{E}},
\end{equation}
which is easily derived from~\eqref{eq:Kumar}. JAP2019 corresponds to the prediction~\eqref{eq:Gsheet}. Both predictions are close, although overestimate the electrical conductance. This deviation may arise due to differences between any particular sample and the imaginary averaged sample. For instance, for any real sample, the angular distribution can be far from the uniform distribution (Fig.~\ref{fig:angles}). Besides, the potential drop along a sample is only \emph{almost} linear rather than \emph{strictly} linear (Fig.~\ref{fig:Vvsxsample4}). In other words, a mean-field approach is based on an assumption that a sample is highly homogeneous and isotropic. This assumption is hardly exact for real-world samples, however, it is expected to be much better for our computer-generated samples, especially for hyperuniform ones.

\begin{figure}[!htb]
  \centering
  \includegraphics[width=\columnwidth]{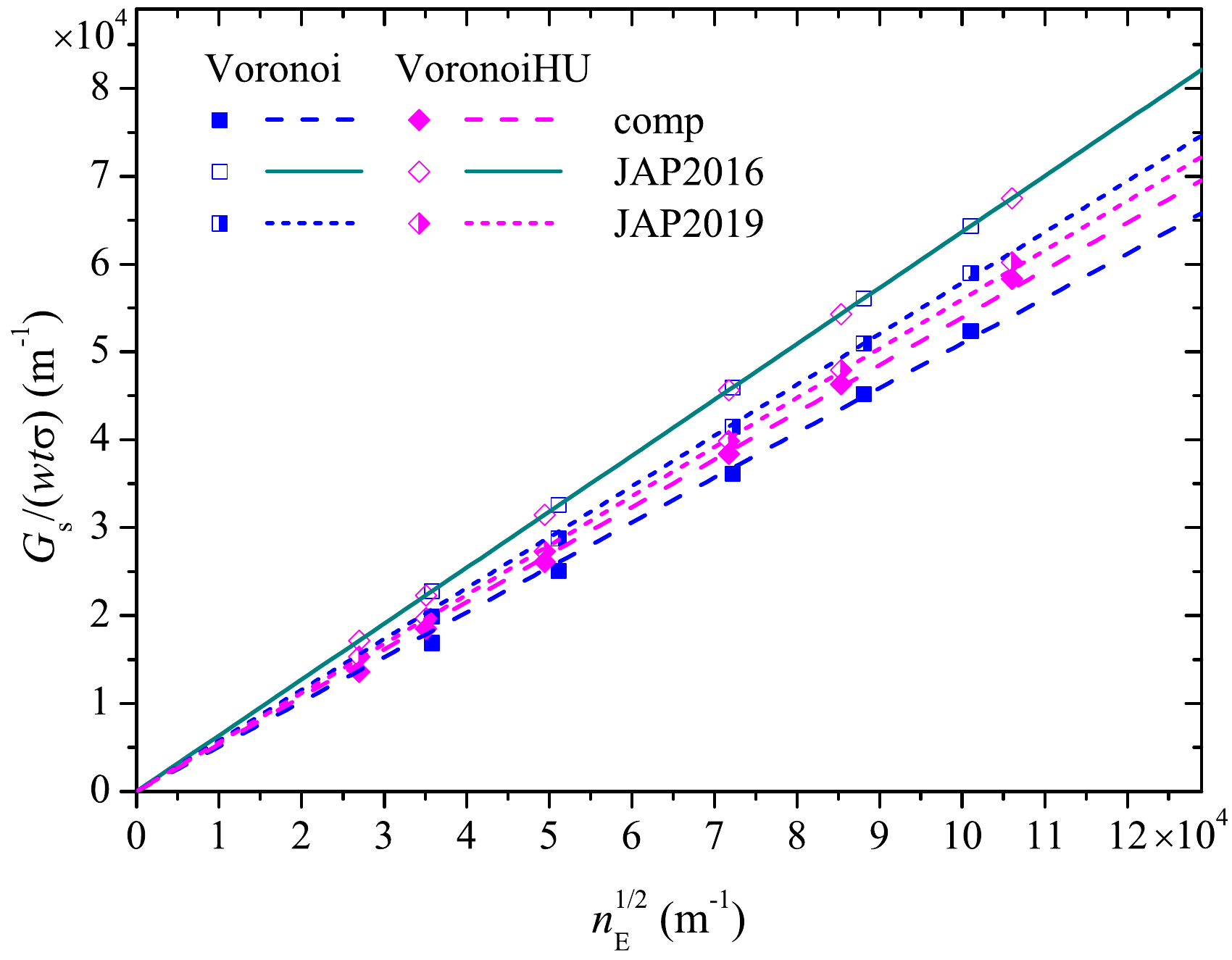}
  \caption{Reduced conductance vs square root of the number density of cracks. JAP2016 corresponds to  prediction~\eqref{eq:GKumar}, JAP2019 corresponds to prediction~\eqref{eq:Gsheet}. Dashed lines correspond to least squares fits.}\label{fig:GsqcomparV}
\end{figure}

Figure~\ref{fig:GsqcomparV} plots the reduced electrical conductance of the computer-generated networks against the square root of the number density of the cracks. It is noticeable that, for hyperuniform samples, prediction~\eqref{eq:Gsheet} is excellent. This suggests that heterogeneity of any particular sample may be a reasonable cause for both~\eqref{eq:Kumar} and~\eqref{eq:Gsheet} to overestimate the electrical conductance. Nevertheless, homogeneity of samples (e.g., transparent heaters) is a natural requirement. Thus, the prediction~\eqref{eq:Gsheet} can be considered as a theoretical limit of the on the maximum electrical conductance that a highly uniform random network can reach.

\section{Conclusion}\label{sec:concl}
We performed image processing and analysis of 10 photos of crack-template based TCFs. The analysis evidenced that (i)~the angle distribution is almost equiprobable; however, the statistical errors are significant, i.e., the properties of a particular sample may differ significantly from the average values (Fig.~\ref{fig:angles}); (ii)~secondary cracks tends to be perpendicular to the primary ones, i.e., T-shaped connections of cracks are dominant; as a result, the typical angles between adjacent cracks are about $90^\circ$ and $180^\circ$ (Fig.~\ref{fig:CrackAnglesBox}); (iii)~the length distribution of crack segments resembles a log-normal distribution (Fig.~\ref{fig:length2p}); (iv)~the average length of crack segments is inversely proportional to the square root of the number density of the cracks  (Fig.~\ref{fig:lmeanvsn}).

We analyzed computer-generated networks obtained using kinds of Voronoi tessellation.
In the first case, the generators (points) were randomly distributed within a square domain. In the second case, discs were deposited in the square domain using random sequential adsorption. The centers of these discs were used as random hyperuniformly distributed points to perform the Voronoi tessellation.
The analysis evidenced that, for both kinds of network, (i)~the angle distribution is equiprobable (Fig.~\ref{fig:angles});  (ii)~Y-shaped connections of cracks dominate; as a result, the typical angles between adjacent cracks are about $120^\circ$ (Fig.~\ref{fig:CrackAnglesBox}); (iii)~the length distribution of the crack segments resembles a log-normal distribution (Fig.~\ref{fig:length2p}); (iv)~the average length of the crack segments is inversely proportional to the square root of the number density of the cracks (Fig.~\ref{fig:lmeanvsn}).

Our computations suggest that (i)~the potential drop along the samples is almost linear  (Fig.~\ref{fig:Vvsxsample4}); (ii)~the theoretically predicted~\cite{Kumar2016,Tarasevich2019} proportionality of the electrical conductance to the square root of the number density of cracks is correct (Fig.~\ref{fig:Gsqcompar}); however, both approaches overestimate the electrical conductance. We suppose, there are two main reason for this overestimation. Both approaches are based largely on the assumptions that (i)~the voltage drop is strictly linear and (ii)~the angle distribution is strictly equiprobable. In fact, both of these requirements are met only approximately. Moreover, approach by \citet{Kumar2016} significantly overestimate the number of cracks intersecting an equipotential line (see Table~\ref{tab:intersect} and Table~\ref{tab:intersectVHU}).

\begin{acknowledgments}
We acknowledge funding from the Russian Science Foundation, Grant No.~23-21-00074. We thank A.S.~Voronin for the unpublished photos of crack templates (denoted as sample~1 and sample~2 in this work).
\end{acknowledgments}

\section*{Data Availability Statement}
The data that support the findings of this study are available from the corresponding author upon reasonable request.

\appendix

\section{Electrical conductance of variable width wire}\label{sec:variablewidth}
Let there be a junction-free random conductive network produced using a crack pattern. Consider one conductive segment between two nearest points of intersection of the cracks. We will assume that this segment is straight and has a length $l$. Let us direct axis $x$ along this conductive segment. We will consider the thickness of the conductor to be a constant $t$, while the width of the conductor is variable $w(x)$.
\begin{figure}[!htb]
  \centering
  \includegraphics[width=0.75\columnwidth]{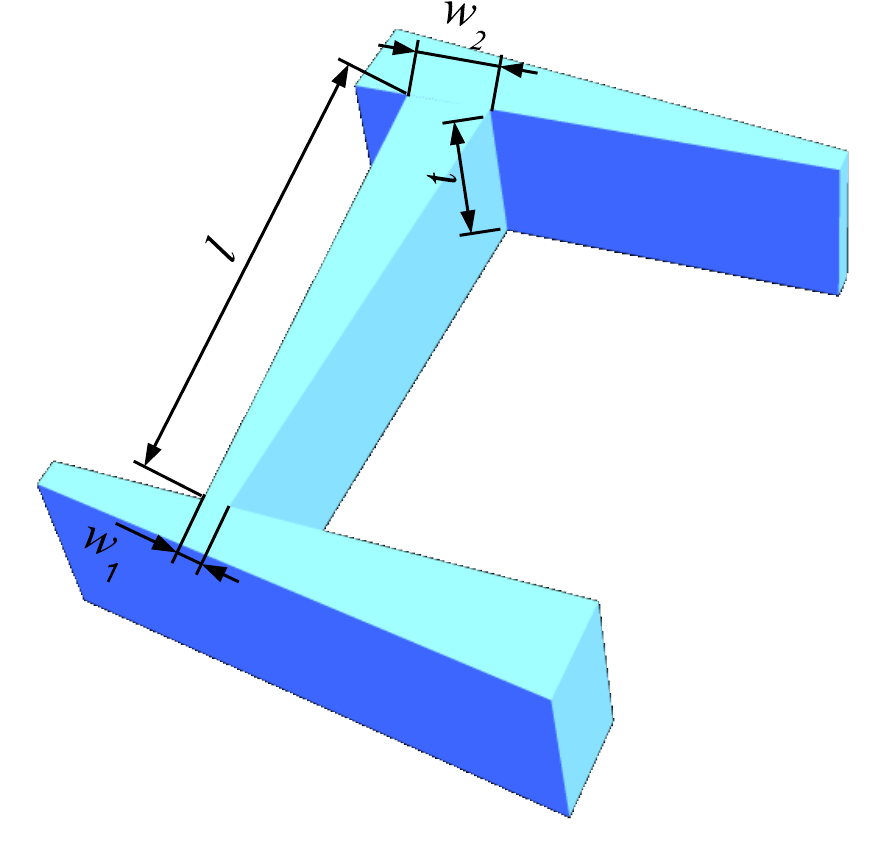}
  \caption{Sizes of a conductive segment.}\label{fig:crack-sizes}
\end{figure}

Let the resistivity of the material filling the cracks be equal to $\rho$. Then the resistance of the segment under consideration is equal to
\begin{equation}\label{eq:Rvariablewidth}
  R = \frac{\rho}{t} \int\limits_{0}^{l} \frac{\mathrm{d}x}{w(x)}.
\end{equation}
Consider the simplest situation, when the width of the conductor changes linearly from $w_1$ to $w_2$ ($w_1 < w_2$)
\begin{equation}\label{eq:variablewidth}
w(x) = w_1 + \frac{w_2 - w_1}{l}x.
\end{equation}
Hence,
\begin{equation}\label{eq:Rvariablewidthint}
R = \frac{\rho l}{t} \int\limits_{0}^{l} \frac{\mathrm{d}x}{ w_1 l + (w_2 - w_1)x} =  \frac{\rho l}{t (w_2 - w_1)} \ln\frac{w_2}{w_1}.
\end{equation}
The electrical conductance of this segment is equal to
\begin{equation}\label{eq:Gvariablewidth}
G  =  \frac{\sigma t}{l}\frac{w_2 - w_1}{\ln w_2 -\ln w_1},
\end{equation}
where $\sigma$ is the electrical conductivity of the material, while the second factor is the logarithmic mean. Introducing a notation $w_2 = \langle w \rangle + \Delta$, $w_1 = \langle w \rangle - \Delta$, we get
\begin{equation}\label{eq:w2w1eval}
\frac{w_2 - w_1}{\ln w_2 -\ln w_1}= \frac{2\Delta}{\ln \frac{\langle w \rangle + \Delta}{\langle w \rangle - \Delta}}.
\end{equation}
When $\Delta \ll \langle w \rangle$,
\begin{equation}\label{eq:lnw2w1}
\ln \frac{\langle w \rangle + \Delta}{\langle w \rangle - \Delta} \approx 2\frac{\Delta}{\langle w \rangle},
\end{equation}
hence,
\begin{equation}\label{eq:lnmeanvariablewidth}
\frac{w_2 - w_1}{\ln w_2 -\ln w_1}\approx \langle w \rangle.
\end{equation}
Thus,
\begin{equation}\label{eq:approxGvariablewidth}
G  \approx  \frac{\sigma t \langle w \rangle}{l}, \quad R \approx  \frac{\rho l}{t \langle w \rangle}.
\end{equation}

\section{Electrical conductance of a crack-template based random network}\label{sec:cond}
Here, we reproduce with minor changes derivations presented in \citet{Tarasevich2019}. These changes are intended to adapt the consideration of a random resistor network produced by randomly placed nanowires  to the case of crack-template-based networks. Consider a network of size $L_x \times L_y$. When the total number of edges (cracks) is $N_\text{E}$, the number density of edges is
\begin{equation}\label{eq:nE}
n_\text{E} = \frac{N_\text{E}}{L_x L_y}.
\end{equation}
Let $l$ be the length of a particular edge, i.e., the length of the crack between the two nearest crossing points (nodes). The resistance of this edge is described by~\eqref{eq:resistor}.

Let a potential difference, $U_0$, be applied to the opposite boundaries of the film along the $x$ direction. Since, in devices such as transparent heaters, solar cells, touch screens, etc., any typical face size (a smallest region of the film bounded by cracks) is much less than the linear film size, a potential drop along the sample is almost linear (Fig.~\ref{fig:Vvsxsample4}). The potential difference between the ends of an $i$-th crack, oriented at angle $\alpha$ with respect to the external electrostatic field,  is
$$
\frac{U_0 l\cos \alpha}{L_x}.
$$
Hence, the electric current in this filled crack is
\begin{equation}\label{eq:i}
i(\alpha) = \frac{U_0 l\cos \alpha}{L_x R} = \frac{U_0 w t \cos \alpha}{L_x \rho }.
\end{equation}

Consider a line, which is perpendicular to the external electrostatic field. A crack, that is oriented with respect to this field at angle $\alpha$, intersects this line, if its origin is located at a distance not exceeding $l \cos \alpha$ from the line. The number of appropriate edges (crack segments) is
$n_\text{E} l L_y \cos \alpha$, while the overall electrical current in all such filled cracks is
$$
\frac{U_0 n_\text{E} l w t L_y\cos^2 \alpha}{\rho L_x}.
$$
The total electric current is
\begin{multline}\label{eq:Itotalsimple}
I = \frac{1}{\pi} \int\limits_{l_\text{min}}^{l_\text{max}}\int\limits_{-\pi/2}^{\pi/2} \frac{U_0 n_\text{E} w t L_y}{\rho L_x}f(l) l\cos^2 \alpha\,\mathrm{d}\alpha\,\mathrm{d}l\\
= \frac{ U_0 n_\text{E} \langle l \rangle w t L_y}{2 \rho L_x}.
\end{multline}
Since the all crack orientations are equiprobable and $\alpha \in [-\pi/2,\pi/2]$, the probability density function (PDF) of the angle orientations is $\pi^{-1}$.  $f(l)$ is the PDF, which describes the distribution of the cracks' length with the distribution's support $l \in [l_\text{min}, l_\text{max}]$. $\langle \cdot \rangle$ denotes the mean value.
Therefore, the electrical sheet conductance is
\begin{equation}\label{eq:Gsheet}
G_\Box = \frac{\sigma n_\text{E}  \langle l \rangle w t}{2},
\end{equation}
where $\sigma = \rho^{-1}$ is the electrical conductivity of the metal. In contrast with~\eqref{eq:GKumar}, the sheet conductance depends not only on the number density of the cracks, but also on the average crack length. However, accounting for  relation~\eqref{eq:meanlvsnE}, the dependency on the average crack length can be eliminated, and as a result the sheet conductance is expected to be dependent on the number density of the cracks as~$\sqrt{n_\text{E}}$.

\section{Relationship between the average length of a crack segment and the concentration of cracks}\label{sec:relationship}

Let a crack network split the sample into $F$ faces (cells). In the case of computer-generated networks based on the Voronoi tessellation, this means using $F$ points for its generation. Let $E$ be the number of edges in this network, and let $\langle l \rangle$ be the average length of one edge. When the sample has dimensions $L_x \times L_y$, the average area of one face is
\begin{equation}\label{eq:meanAF}
\langle A_\text{F} \rangle = \frac{L_x  L_y}{F},
\end{equation}
while the number density of the edges is defined by~\eqref{eq:nE}.
Obviously,
\begin{equation}\label{eq:degV}
\sum\limits_{i=1}^{V} \deg V_i = 2E
\end{equation}
or
\begin{equation}\label{eq:meandegV}
\frac{1}{V}\sum\limits_{i=1}^{V} \deg V_i = \frac{2E}{V},
\quad
\langle \deg V \rangle = \frac{2E}{V},
\quad
V = \frac{2E}{\langle \deg V \rangle}.
\end{equation}
According to Euler's formula for planar graphs (networks) $V - E + F = 2$, where $V$ is the number of vertices, while the number of faces, $F$, accounts for the outer, infinitely large face. Since the outer face is out of our interest, $F = E - V + 1$.
Therefore,
\begin{equation}\label{eq:F}
F = \left( 1 - \frac{2}{\langle \deg V \rangle}\right) E + 1.
\end{equation}
Note, that if the number of faces is large ($F \gg 1$), then $ E - V \approx F$.
Let's move from the number of edges, $E$, to their number density~\eqref{eq:nE} and the average face area~\eqref{eq:meanAF}
\begin{equation}\label{eq:invAfmean}
\langle A_\text{F} \rangle^{-1} = \left( 1 - \frac{2}{\langle \deg V \rangle}\right) n_\text{E} + \frac{1}{L_x L_y}.
\end{equation}
When $E \gg 1$, which is ensured in the case of crack patterns,
\begin{equation}\label{eq:invmeanAF}
\langle A_\text{F} \rangle^{-1} \approx \left( 1 - \frac{2}{\langle \deg V \rangle}\right) n_\text{E}.
\end{equation}
Since, in the case of computer-generated networks based on the Voronoi tessellation, $\langle \deg V \rangle \approx 3$,
\begin{equation}\label{eq:estimate}
\langle A_\text{F} \rangle^{-1} \approx\frac{n_\text{E}}{3}.
\end{equation}

A circle has the smallest perimeter among all figures of the same area, hence, the perimeter of one face is limited by the value of the circumference, determined as follows. Since, for a circle,
$\langle A_\text{F} \rangle = \pi r^2$, then
\begin{equation}\label{eq:rcirc}
r = \sqrt{\frac{\langle A_\text{F} \rangle }{\pi}}.
\end{equation}
Whereas the circumference is $C =2 \pi r$, then
\begin{equation}\label{eq:meanCcirc}
\langle C \rangle = 2 \pi \sqrt{\frac{\langle A_\text{F} \rangle }{\pi}} =  2 \sqrt{\langle A_\text{F} \rangle \pi}.
\end{equation}
The perimeter of any face is obviously greater than the circumference of a circle of the same area.

Since each edge belongs to two faces, the total length of all edges, $\langle l \rangle E$, is greater or equal to $F\langle C \rangle /2$, i.e.,
$\langle l \rangle E \geqslant F \sqrt{\pi\langle A_\text{F} \rangle}.$
Using the approximate value of the average face area~\eqref{eq:estimate},
\begin{equation}\label{eq:minlvsnE}
\langle l \rangle  \gtrapprox \sqrt{ \frac{\pi}{3 n_\text{E}}}  \approx 1.023 n_\text{E}^{-1/2 }.
\end{equation}
Hence,
\begin{equation}\label{eq:circle}
 n_\text{E} \langle l \rangle\gtrapprox \sqrt{ \frac{\pi n_\text{E}}{3 }}.
\end{equation}

If the domain is divided into identical square cells, then the number of such cells (faces) is equal to
\begin{equation}\label{eq:Fsquare}
F = \frac{L_x L_y}{\langle l \rangle^2}.
\end{equation}
The number of edges equals
\begin{equation}\label{eq:Esquare}
E = 2F = \frac{2 L_x L_y}{\langle l \rangle^2}.
\end{equation}
The number density of edges is equal to
\begin{equation}\label{eq:nEsquare}
n_\text{E} = \frac{2}{\langle l \rangle^2}.
\end{equation}
Hence, the average edge length depends on the number density of the cracks as follows
\begin{equation}\label{eq:meanlsquare}
\langle l \rangle = \sqrt{ \frac{2}{n_\text{E}} }.
\end{equation}
The total edge length is
\begin{equation}\label{eq:totallsquare}
\frac{2 \langle l \rangle L_x L_y}{\langle l \rangle^2}.
\end{equation}
The  area of one face (cell) equals
\begin{equation}\label{eq:AFsquare}
A_F  = \langle l \rangle^2.
\end{equation}

If the partitioning of the area is made in the form of a brickwork pattern, then the degrees of the vertices are 3, and the angles between the edges are straight, which resembles the properties of crack networks.
The number of faces (cells) is equal to
\begin{equation}\label{eq:Fbrick}
F = \frac{L_x L_y}{2 \langle l \rangle^2}.
\end{equation}
The number of edges equals
\begin{equation}\label{eq:Ebrick}
E = 3F = \frac{3 L_x L_y}{2 \langle l \rangle^2}.
\end{equation}
The number density of the edges is
\begin{equation}\label{eq:nEbrick}
n_\text{E} = \frac{3}{2 \langle l \rangle^2}.
\end{equation}
The average edge length depends on the number density of edges as follows
\begin{equation}\label{eq:lbrick}
\langle l \rangle = \sqrt{ \frac{3}{2 n_\text{E}} }\approx 1.225  n_\text{E}^{-1/2}.
\end{equation}
Hence,
\begin{equation}\label{eq:brick}
  n_\text{E} \langle l \rangle = \sqrt{ \frac{3 n_\text{E}}{2} }.
\end{equation}

\bibliography{cracksR1}

\begin{thebibliography}{52}%
\makeatletter
\providecommand \@ifxundefined [1]{%
 \@ifx{#1\undefined}
}%
\providecommand \@ifnum [1]{%
 \ifnum #1\expandafter \@firstoftwo
 \else \expandafter \@secondoftwo
 \fi
}%
\providecommand \@ifx [1]{%
 \ifx #1\expandafter \@firstoftwo
 \else \expandafter \@secondoftwo
 \fi
}%
\providecommand \natexlab [1]{#1}%
\providecommand \enquote  [1]{``#1''}%
\providecommand \bibnamefont  [1]{#1}%
\providecommand \bibfnamefont [1]{#1}%
\providecommand \citenamefont [1]{#1}%
\providecommand \href@noop [0]{\@secondoftwo}%
\providecommand \href [0]{\begingroup \@sanitize@url \@href}%
\providecommand \@href[1]{\@@startlink{#1}\@@href}%
\providecommand \@@href[1]{\endgroup#1\@@endlink}%
\providecommand \@sanitize@url [0]{\catcode `\\12\catcode `\$12\catcode
  `\&12\catcode `\#12\catcode `\^12\catcode `\_12\catcode `\%12\relax}%
\providecommand \@@startlink[1]{}%
\providecommand \@@endlink[0]{}%
\providecommand \url  [0]{\begingroup\@sanitize@url \@url }%
\providecommand \@url [1]{\endgroup\@href {#1}{\urlprefix }}%
\providecommand \urlprefix  [0]{URL }%
\providecommand \Eprint [0]{\href }%
\providecommand \doibase [0]{http://dx.doi.org/}%
\providecommand \selectlanguage [0]{\@gobble}%
\providecommand \bibinfo  [0]{\@secondoftwo}%
\providecommand \bibfield  [0]{\@secondoftwo}%
\providecommand \translation [1]{[#1]}%
\providecommand \BibitemOpen [0]{}%
\providecommand \bibitemStop [0]{}%
\providecommand \bibitemNoStop [0]{.\EOS\space}%
\providecommand \EOS [0]{\spacefactor3000\relax}%
\providecommand \BibitemShut  [1]{\csname bibitem#1\endcsname}%
\let\auto@bib@innerbib\@empty
\bibitem [{\citenamefont {Liu}\ \emph {et~al.}(2016)\citenamefont {Liu},
  \citenamefont {Shen}, \citenamefont {Hu},\ and\ \citenamefont
  {Chen}}]{Liu2016}%
  \BibitemOpen
  \bibfield  {author} {\bibinfo {author} {\bibfnamefont {Yanhua}\ \bibnamefont
  {Liu}}, \bibinfo {author} {\bibfnamefont {Su}~\bibnamefont {Shen}}, \bibinfo
  {author} {\bibfnamefont {Jin}\ \bibnamefont {Hu}}, \ and\ \bibinfo {author}
  {\bibfnamefont {Linsen}\ \bibnamefont {Chen}},\ }\bibfield  {title} {\enquote
  {\bibinfo {title} {Embedded {Ag} mesh electrodes for polymer dispersed liquid
  crystal devices on flexible substrate},}\ }\href {\doibase
  10.1364/oe.24.025774} {\bibfield  {journal} {\bibinfo  {journal} {Opt.
  Express}\ }\textbf {\bibinfo {volume} {24}},\ \bibinfo {pages} {25774}
  (\bibinfo {year} {2016})}\BibitemShut {NoStop}%
\bibitem [{\citenamefont {Han}\ \emph {et~al.}(2016)\citenamefont {Han},
  \citenamefont {Lin}, \citenamefont {Liu}, \citenamefont {Fu}, \citenamefont
  {Ma}, \citenamefont {Jin},\ and\ \citenamefont {Tan}}]{Han2016}%
  \BibitemOpen
  \bibfield  {author} {\bibinfo {author} {\bibfnamefont {Yu}~\bibnamefont
  {Han}}, \bibinfo {author} {\bibfnamefont {Jie}\ \bibnamefont {Lin}}, \bibinfo
  {author} {\bibfnamefont {Yuxuan}\ \bibnamefont {Liu}}, \bibinfo {author}
  {\bibfnamefont {Hao}\ \bibnamefont {Fu}}, \bibinfo {author} {\bibfnamefont
  {Yuan}\ \bibnamefont {Ma}}, \bibinfo {author} {\bibfnamefont {Peng}\
  \bibnamefont {Jin}}, \ and\ \bibinfo {author} {\bibfnamefont {Jiubin}\
  \bibnamefont {Tan}},\ }\bibfield  {title} {\enquote {\bibinfo {title}
  {Crackle template based metallic mesh with highly homogeneous light
  transmission for high-performance transparent {EMI} shielding},}\ }\href
  {\doibase 10.1038/srep25601} {\bibfield  {journal} {\bibinfo  {journal} {Sci.
  Rep.}\ }\textbf {\bibinfo {volume} {6}},\ \bibinfo {pages} {25601} (\bibinfo
  {year} {2016})}\BibitemShut {NoStop}%
\bibitem [{\citenamefont {Shen}\ \emph {et~al.}(2018)\citenamefont {Shen},
  \citenamefont {Chen}, \citenamefont {Zhang},\ and\ \citenamefont
  {Liu}}]{Shen2018}%
  \BibitemOpen
  \bibfield  {author} {\bibinfo {author} {\bibfnamefont {Su}~\bibnamefont
  {Shen}}, \bibinfo {author} {\bibfnamefont {Shi-Yu}\ \bibnamefont {Chen}},
  \bibinfo {author} {\bibfnamefont {Dong-Yu}\ \bibnamefont {Zhang}}, \ and\
  \bibinfo {author} {\bibfnamefont {Yan-Hua}\ \bibnamefont {Liu}},\ }\bibfield
  {title} {\enquote {\bibinfo {title} {High-performance composite {Ag}--{Ni}
  mesh based flexible transparent conductive film as multifunctional
  devices},}\ }\href {\doibase 10.1364/oe.26.027545} {\bibfield  {journal}
  {\bibinfo  {journal} {Opt. Express}\ }\textbf {\bibinfo {volume} {26}},\
  \bibinfo {pages} {27545} (\bibinfo {year} {2018})}\BibitemShut {NoStop}%
\bibitem [{\citenamefont {Jiang}\ \emph {et~al.}(2019)\citenamefont {Jiang},
  \citenamefont {Huang}, \citenamefont {Chen},\ and\ \citenamefont
  {Liu}}]{Jiang2019}%
  \BibitemOpen
  \bibfield  {author} {\bibinfo {author} {\bibfnamefont {Zhou-ying}\
  \bibnamefont {Jiang}}, \bibinfo {author} {\bibfnamefont {Wenbin}\
  \bibnamefont {Huang}}, \bibinfo {author} {\bibfnamefont {Lin-sen}\
  \bibnamefont {Chen}}, \ and\ \bibinfo {author} {\bibfnamefont {Yan-hua}\
  \bibnamefont {Liu}},\ }\bibfield  {title} {\enquote {\bibinfo {title}
  {Ultrathin, lightweight, and freestanding metallic mesh for transparent
  electromagnetic interference shielding},}\ }\href {\doibase
  10.1364/oe.27.024194} {\bibfield  {journal} {\bibinfo  {journal} {Opt.
  Express}\ }\textbf {\bibinfo {volume} {27}},\ \bibinfo {pages} {24194}
  (\bibinfo {year} {2019})}\BibitemShut {NoStop}%
\bibitem [{\citenamefont {Song}\ \emph {et~al.}(2023)\citenamefont {Song},
  \citenamefont {Sun}, \citenamefont {Xu}, \citenamefont {Shan}, \citenamefont
  {Tang}, \citenamefont {Tian}, \citenamefont {Xu},\ and\ \citenamefont
  {Gao}}]{Song2023}%
  \BibitemOpen
  \bibfield  {author} {\bibinfo {author} {\bibfnamefont {Naitao}\ \bibnamefont
  {Song}}, \bibinfo {author} {\bibfnamefont {Qiao}\ \bibnamefont {Sun}},
  \bibinfo {author} {\bibfnamefont {Su}~\bibnamefont {Xu}}, \bibinfo {author}
  {\bibfnamefont {Dongzhi}\ \bibnamefont {Shan}}, \bibinfo {author}
  {\bibfnamefont {Yang}\ \bibnamefont {Tang}}, \bibinfo {author} {\bibfnamefont
  {Xiaoxi}\ \bibnamefont {Tian}}, \bibinfo {author} {\bibfnamefont {Nianxi}\
  \bibnamefont {Xu}}, \ and\ \bibinfo {author} {\bibfnamefont {Jinsong}\
  \bibnamefont {Gao}},\ }\bibfield  {title} {\enquote {\bibinfo {title}
  {Ultra-wide-band optically transparent anti-diffraction metamaterial absorber
  with a thiessen-polygon metal-mesh shielding layer},}\ }\href {\doibase
  10.1364/prj.486613} {\bibfield  {journal} {\bibinfo  {journal} {Photonics
  Res.}\ } (\bibinfo {year} {2023}),\ 10.1364/prj.486613}\BibitemShut {NoStop}%
\bibitem [{\citenamefont {Li}\ \emph {et~al.}(2023)\citenamefont {Li},
  \citenamefont {Zi}, \citenamefont {Zhu}, \citenamefont {Zhang}, \citenamefont
  {Tai}, \citenamefont {Wang}, \citenamefont {Sun}, \citenamefont {Zhang},
  \citenamefont {Ge}, \citenamefont {Huang}, \citenamefont {Liu}, \citenamefont
  {Yang}, \citenamefont {Yang},\ and\ \citenamefont {Lan}}]{Li2023}%
  \BibitemOpen
  \bibfield  {author} {\bibinfo {author} {\bibfnamefont {Hongke}\ \bibnamefont
  {Li}}, \bibinfo {author} {\bibfnamefont {Denghua}\ \bibnamefont {Zi}},
  \bibinfo {author} {\bibfnamefont {Xiaoyang}\ \bibnamefont {Zhu}}, \bibinfo
  {author} {\bibfnamefont {Houchao}\ \bibnamefont {Zhang}}, \bibinfo {author}
  {\bibfnamefont {Yuping}\ \bibnamefont {Tai}}, \bibinfo {author}
  {\bibfnamefont {Rui}\ \bibnamefont {Wang}}, \bibinfo {author} {\bibfnamefont
  {Luanfa}\ \bibnamefont {Sun}}, \bibinfo {author} {\bibfnamefont {Youchao}\
  \bibnamefont {Zhang}}, \bibinfo {author} {\bibfnamefont {Wensong}\
  \bibnamefont {Ge}}, \bibinfo {author} {\bibfnamefont {Youqi}\ \bibnamefont
  {Huang}}, \bibinfo {author} {\bibfnamefont {Gang}\ \bibnamefont {Liu}},
  \bibinfo {author} {\bibfnamefont {Wenchao}\ \bibnamefont {Yang}}, \bibinfo
  {author} {\bibfnamefont {Jianjun}\ \bibnamefont {Yang}}, \ and\ \bibinfo
  {author} {\bibfnamefont {Hongbo}\ \bibnamefont {Lan}},\ }\bibfield  {title}
  {\enquote {\bibinfo {title} {Electric field driven printing of repeatable
  random metal meshes for flexible transparent electrodes},}\ }\href {\doibase
  10.1016/j.optlastec.2022.108730} {\bibfield  {journal} {\bibinfo  {journal}
  {Opt. Laser Technol.}\ }\textbf {\bibinfo {volume} {157}},\ \bibinfo {pages}
  {108730} (\bibinfo {year} {2023})}\BibitemShut {NoStop}%
\bibitem [{\citenamefont {Suh}\ \emph {et~al.}(2016{\natexlab{a}})\citenamefont
  {Suh}, \citenamefont {Kwon}, \citenamefont {Lee}, \citenamefont {Lee},
  \citenamefont {Jeong}, \citenamefont {Kim}, \citenamefont {Cho},
  \citenamefont {Yeo},\ and\ \citenamefont {Ko}}]{Suh2016}%
  \BibitemOpen
  \bibfield  {author} {\bibinfo {author} {\bibfnamefont {Young~D.}\
  \bibnamefont {Suh}}, \bibinfo {author} {\bibfnamefont {Jinhyeong}\
  \bibnamefont {Kwon}}, \bibinfo {author} {\bibfnamefont {Jinhwan}\
  \bibnamefont {Lee}}, \bibinfo {author} {\bibfnamefont {Habeom}\ \bibnamefont
  {Lee}}, \bibinfo {author} {\bibfnamefont {Seongmin}\ \bibnamefont {Jeong}},
  \bibinfo {author} {\bibfnamefont {Dongkwan}\ \bibnamefont {Kim}}, \bibinfo
  {author} {\bibfnamefont {Hyunmin}\ \bibnamefont {Cho}}, \bibinfo {author}
  {\bibfnamefont {Junyeob}\ \bibnamefont {Yeo}}, \ and\ \bibinfo {author}
  {\bibfnamefont {Seung~Hwan}\ \bibnamefont {Ko}},\ }\bibfield  {title}
  {\enquote {\bibinfo {title} {Maskless fabrication of highly robust, flexible
  transparent {Cu} conductor by random crack network assisted {Cu} nanoparticle
  patterning and laser sintering},}\ }\href {\doibase 10.1002/aelm.201600277}
  {\bibfield  {journal} {\bibinfo  {journal} {Adv. Electron. Mater.}\ }\textbf
  {\bibinfo {volume} {2}},\ \bibinfo {pages} {1600277} (\bibinfo {year}
  {2016}{\natexlab{a}})}\BibitemShut {NoStop}%
\bibitem [{\citenamefont {Suh}\ \emph {et~al.}(2016{\natexlab{b}})\citenamefont
  {Suh}, \citenamefont {Hong}, \citenamefont {Lee}, \citenamefont {Lee},
  \citenamefont {Jung}, \citenamefont {Kwon}, \citenamefont {Moon},
  \citenamefont {Won}, \citenamefont {Shin}, \citenamefont {Yeo},\ and\
  \citenamefont {Ko}}]{Suh2016a}%
  \BibitemOpen
  \bibfield  {author} {\bibinfo {author} {\bibfnamefont {Young~D.}\
  \bibnamefont {Suh}}, \bibinfo {author} {\bibfnamefont {Sukjoon}\ \bibnamefont
  {Hong}}, \bibinfo {author} {\bibfnamefont {Jinhwan}\ \bibnamefont {Lee}},
  \bibinfo {author} {\bibfnamefont {Habeom}\ \bibnamefont {Lee}}, \bibinfo
  {author} {\bibfnamefont {Seongmin}\ \bibnamefont {Jung}}, \bibinfo {author}
  {\bibfnamefont {Jinhyeong}\ \bibnamefont {Kwon}}, \bibinfo {author}
  {\bibfnamefont {Hyunjin}\ \bibnamefont {Moon}}, \bibinfo {author}
  {\bibfnamefont {Phillip}\ \bibnamefont {Won}}, \bibinfo {author}
  {\bibfnamefont {Jaeho}\ \bibnamefont {Shin}}, \bibinfo {author}
  {\bibfnamefont {Junyeob}\ \bibnamefont {Yeo}}, \ and\ \bibinfo {author}
  {\bibfnamefont {Seung~Hwan}\ \bibnamefont {Ko}},\ }\bibfield  {title}
  {\enquote {\bibinfo {title} {Random nanocrack, assisted metal
  nanowire-bundled network fabrication for a highly flexible and transparent
  conductor},}\ }\href {\doibase 10.1039/c6ra11467a} {\bibfield  {journal}
  {\bibinfo  {journal} {RSC Adv.}\ }\textbf {\bibinfo {volume} {6}},\ \bibinfo
  {pages} {57434--57440} (\bibinfo {year} {2016}{\natexlab{b}})}\BibitemShut
  {NoStop}%
\bibitem [{\citenamefont {Jung}\ \emph {et~al.}(2019)\citenamefont {Jung},
  \citenamefont {Cho}, \citenamefont {Choi}, \citenamefont {Kim}, \citenamefont
  {Kwon}, \citenamefont {Shin}, \citenamefont {Hong}, \citenamefont {Kim},
  \citenamefont {Yoon}, \citenamefont {Lee}, \citenamefont {Lee}, \citenamefont
  {Suh},\ and\ \citenamefont {Ko}}]{Jung2019}%
  \BibitemOpen
  \bibfield  {author} {\bibinfo {author} {\bibfnamefont {Jinwook}\ \bibnamefont
  {Jung}}, \bibinfo {author} {\bibfnamefont {Hyunmin}\ \bibnamefont {Cho}},
  \bibinfo {author} {\bibfnamefont {Seok~Hwan}\ \bibnamefont {Choi}}, \bibinfo
  {author} {\bibfnamefont {Dongkwan}\ \bibnamefont {Kim}}, \bibinfo {author}
  {\bibfnamefont {Jinhyeong}\ \bibnamefont {Kwon}}, \bibinfo {author}
  {\bibfnamefont {Jaeho}\ \bibnamefont {Shin}}, \bibinfo {author}
  {\bibfnamefont {Sukjoon}\ \bibnamefont {Hong}}, \bibinfo {author}
  {\bibfnamefont {Hyeonseok}\ \bibnamefont {Kim}}, \bibinfo {author}
  {\bibfnamefont {Yeosang}\ \bibnamefont {Yoon}}, \bibinfo {author}
  {\bibfnamefont {Jinwoo}\ \bibnamefont {Lee}}, \bibinfo {author}
  {\bibfnamefont {Daeho}\ \bibnamefont {Lee}}, \bibinfo {author} {\bibfnamefont
  {Young~D.}\ \bibnamefont {Suh}}, \ and\ \bibinfo {author} {\bibfnamefont
  {Seung~Hwan}\ \bibnamefont {Ko}},\ }\bibfield  {title} {\enquote {\bibinfo
  {title} {Moir{\'{e}}-free imperceptible and flexible random metal grid
  electrodes with large figure-of-merit by photonic sintering control of copper
  nanoparticles},}\ }\href {\doibase 10.1021/acsami.9b01893} {\bibfield
  {journal} {\bibinfo  {journal} {ACS Appl. Mater. Interfaces}\ }\textbf
  {\bibinfo {volume} {11}},\ \bibinfo {pages} {15773--15780} (\bibinfo {year}
  {2019})}\BibitemShut {NoStop}%
\bibitem [{\citenamefont {Walia}\ \emph {et~al.}(2019)\citenamefont {Walia},
  \citenamefont {Mondal},\ and\ \citenamefont {Kulkarni}}]{Walia2019}%
  \BibitemOpen
  \bibfield  {author} {\bibinfo {author} {\bibfnamefont {Sunil}\ \bibnamefont
  {Walia}}, \bibinfo {author} {\bibfnamefont {Indrajit}\ \bibnamefont
  {Mondal}}, \ and\ \bibinfo {author} {\bibfnamefont {Giridhar~U.}\
  \bibnamefont {Kulkarni}},\ }\bibfield  {title} {\enquote {\bibinfo {title}
  {Patterned {Cu}-mesh-based transparent and wearable touch panel for tactile,
  proximity, pressure, and temperature sensing},}\ }\href {\doibase
  10.1021/acsaelm.9b00330} {\bibfield  {journal} {\bibinfo  {journal} {ACS
  Appl. Electron. Mater.}\ }\textbf {\bibinfo {volume} {1}},\ \bibinfo {pages}
  {1597--1604} (\bibinfo {year} {2019})}\BibitemShut {NoStop}%
\bibitem [{\citenamefont {Melnychenko}\ and\ \citenamefont
  {Kudrawiec}(2022)}]{Melnychenko2022}%
  \BibitemOpen
  \bibfield  {author} {\bibinfo {author} {\bibfnamefont {Anna~M.}\ \bibnamefont
  {Melnychenko}}\ and\ \bibinfo {author} {\bibfnamefont {Robert}\ \bibnamefont
  {Kudrawiec}},\ }\bibfield  {title} {\enquote {\bibinfo {title}
  {Crack-templated wire-like semitransparent electrodes with unique irregular
  patterns},}\ }\href {\doibase 10.1021/acsomega.2c05131} {\bibfield  {journal}
  {\bibinfo  {journal} {ACS Omega}\ }\textbf {\bibinfo {volume} {7}},\ \bibinfo
  {pages} {39181--39186} (\bibinfo {year} {2022})}\BibitemShut {NoStop}%
\bibitem [{\citenamefont {Lee}\ \emph {et~al.}(2019)\citenamefont {Lee},
  \citenamefont {Jin}, \citenamefont {Ovhal}, \citenamefont {Kumar},\ and\
  \citenamefont {Kang}}]{Lee2019}%
  \BibitemOpen
  \bibfield  {author} {\bibinfo {author} {\bibfnamefont {Hock~Beng}\
  \bibnamefont {Lee}}, \bibinfo {author} {\bibfnamefont {Won-Yong}\
  \bibnamefont {Jin}}, \bibinfo {author} {\bibfnamefont {Manoj~Mayaji}\
  \bibnamefont {Ovhal}}, \bibinfo {author} {\bibfnamefont {Neetesh}\
  \bibnamefont {Kumar}}, \ and\ \bibinfo {author} {\bibfnamefont {Jae-Wook}\
  \bibnamefont {Kang}},\ }\bibfield  {title} {\enquote {\bibinfo {title}
  {Flexible transparent conducting electrodes based on metal meshes for organic
  optoelectronic device applications: a review},}\ }\href {\doibase
  10.1039/c8tc04423f} {\bibfield  {journal} {\bibinfo  {journal} {J. Mater.
  Chem. C}\ }\textbf {\bibinfo {volume} {7}},\ \bibinfo {pages} {1087--1110}
  (\bibinfo {year} {2019})}\BibitemShut {NoStop}%
\bibitem [{\citenamefont {Rao}\ and\ \citenamefont
  {Kulkarni}(2014)}]{Rao2014a}%
  \BibitemOpen
  \bibfield  {author} {\bibinfo {author} {\bibfnamefont {K.~D.~M.}\
  \bibnamefont {Rao}}\ and\ \bibinfo {author} {\bibfnamefont {Giridhar~U.}\
  \bibnamefont {Kulkarni}},\ }\bibfield  {title} {\enquote {\bibinfo {title} {A
  highly crystalline single {Au} wire network as a high temperature transparent
  heater},}\ }\href {\doibase 10.1039/c4nr00869c} {\bibfield  {journal}
  {\bibinfo  {journal} {Nanoscale}\ }\textbf {\bibinfo {volume} {6}},\ \bibinfo
  {pages} {5645} (\bibinfo {year} {2014})}\BibitemShut {NoStop}%
\bibitem [{\citenamefont {Han}\ \emph {et~al.}(2014)\citenamefont {Han},
  \citenamefont {Pei}, \citenamefont {Huang}, \citenamefont {Zhang},
  \citenamefont {Rong}, \citenamefont {Lin}, \citenamefont {Guo}, \citenamefont
  {Sun}, \citenamefont {Guo}, \citenamefont {Carnahan}, \citenamefont
  {Giersig}, \citenamefont {Wang}, \citenamefont {Gao}, \citenamefont {Ren},\
  and\ \citenamefont {Kempa}}]{Han2014}%
  \BibitemOpen
  \bibfield  {author} {\bibinfo {author} {\bibfnamefont {Bing}\ \bibnamefont
  {Han}}, \bibinfo {author} {\bibfnamefont {Ke}~\bibnamefont {Pei}}, \bibinfo
  {author} {\bibfnamefont {Yuanlin}\ \bibnamefont {Huang}}, \bibinfo {author}
  {\bibfnamefont {Xiaojian}\ \bibnamefont {Zhang}}, \bibinfo {author}
  {\bibfnamefont {Qikun}\ \bibnamefont {Rong}}, \bibinfo {author}
  {\bibfnamefont {Qinggeng}\ \bibnamefont {Lin}}, \bibinfo {author}
  {\bibfnamefont {Yangfei}\ \bibnamefont {Guo}}, \bibinfo {author}
  {\bibfnamefont {Tianyi}\ \bibnamefont {Sun}}, \bibinfo {author}
  {\bibfnamefont {Chuanfei}\ \bibnamefont {Guo}}, \bibinfo {author}
  {\bibfnamefont {David}\ \bibnamefont {Carnahan}}, \bibinfo {author}
  {\bibfnamefont {Michael}\ \bibnamefont {Giersig}}, \bibinfo {author}
  {\bibfnamefont {Yang}\ \bibnamefont {Wang}}, \bibinfo {author} {\bibfnamefont
  {Jinwei}\ \bibnamefont {Gao}}, \bibinfo {author} {\bibfnamefont {Zhifeng}\
  \bibnamefont {Ren}}, \ and\ \bibinfo {author} {\bibfnamefont {Krzysztof}\
  \bibnamefont {Kempa}},\ }\bibfield  {title} {\enquote {\bibinfo {title}
  {Uniform self-forming metallic network as a high-performance transparent
  conductive electrode},}\ }\href {\doibase 10.1002/adma.201302950} {\bibfield
  {journal} {\bibinfo  {journal} {Adv. Mater.}\ }\textbf {\bibinfo {volume}
  {26}},\ \bibinfo {pages} {873--877} (\bibinfo {year} {2014})}\BibitemShut
  {NoStop}%
\bibitem [{\citenamefont {Rao}\ \emph {et~al.}(2014)\citenamefont {Rao},
  \citenamefont {Hunger}, \citenamefont {Gupta}, \citenamefont {Kulkarni},\
  and\ \citenamefont {Thelakkat}}]{Rao2014}%
  \BibitemOpen
  \bibfield  {author} {\bibinfo {author} {\bibfnamefont {K.~D.~M.}\
  \bibnamefont {Rao}}, \bibinfo {author} {\bibfnamefont {Christoph}\
  \bibnamefont {Hunger}}, \bibinfo {author} {\bibfnamefont {Ritu}\ \bibnamefont
  {Gupta}}, \bibinfo {author} {\bibfnamefont {Giridhar~U.}\ \bibnamefont
  {Kulkarni}}, \ and\ \bibinfo {author} {\bibfnamefont {Mukundan}\ \bibnamefont
  {Thelakkat}},\ }\bibfield  {title} {\enquote {\bibinfo {title} {A cracked
  polymer templated metal network as a transparent conducting electrode for
  {ITO}-free organic solar cells},}\ }\href {\doibase 10.1039/c4cp02250e}
  {\bibfield  {journal} {\bibinfo  {journal} {Phys. Chem. Chem. Phys.}\
  }\textbf {\bibinfo {volume} {16}},\ \bibinfo {pages} {15107--15110} (\bibinfo
  {year} {2014})}\BibitemShut {NoStop}%
\bibitem [{\citenamefont {Pei}\ \emph {et~al.}(2015)\citenamefont {Pei},
  \citenamefont {Wang}, \citenamefont {Ren}, \citenamefont {Zhang},
  \citenamefont {Peng},\ and\ \citenamefont {Chan}}]{Pei2015}%
  \BibitemOpen
  \bibfield  {author} {\bibinfo {author} {\bibfnamefont {Ke}~\bibnamefont
  {Pei}}, \bibinfo {author} {\bibfnamefont {Zongrong}\ \bibnamefont {Wang}},
  \bibinfo {author} {\bibfnamefont {Xiaochen}\ \bibnamefont {Ren}}, \bibinfo
  {author} {\bibfnamefont {Zhichao}\ \bibnamefont {Zhang}}, \bibinfo {author}
  {\bibfnamefont {Boyu}\ \bibnamefont {Peng}}, \ and\ \bibinfo {author}
  {\bibfnamefont {Paddy K.~L.}\ \bibnamefont {Chan}},\ }\bibfield  {title}
  {\enquote {\bibinfo {title} {Fully transparent organic transistors with
  junction-free metallic network electrodes},}\ }\href {\doibase
  10.1063/1.4927445} {\bibfield  {journal} {\bibinfo  {journal} {Appl. Phys.
  Lett.}\ }\textbf {\bibinfo {volume} {107}},\ \bibinfo {pages} {033302}
  (\bibinfo {year} {2015})}\BibitemShut {NoStop}%
\bibitem [{\citenamefont {Peng}\ \emph {et~al.}(2016)\citenamefont {Peng},
  \citenamefont {Li}, \citenamefont {Han}, \citenamefont {Rong}, \citenamefont
  {Lu}, \citenamefont {Wang}, \citenamefont {Zeng}, \citenamefont {Zhou},
  \citenamefont {Liu}, \citenamefont {Kempa},\ and\ \citenamefont
  {Gao}}]{Peng2016}%
  \BibitemOpen
  \bibfield  {author} {\bibinfo {author} {\bibfnamefont {Qiang}\ \bibnamefont
  {Peng}}, \bibinfo {author} {\bibfnamefont {Songru}\ \bibnamefont {Li}},
  \bibinfo {author} {\bibfnamefont {Bing}\ \bibnamefont {Han}}, \bibinfo
  {author} {\bibfnamefont {Qikun}\ \bibnamefont {Rong}}, \bibinfo {author}
  {\bibfnamefont {Xubing}\ \bibnamefont {Lu}}, \bibinfo {author} {\bibfnamefont
  {Qianming}\ \bibnamefont {Wang}}, \bibinfo {author} {\bibfnamefont {Min}\
  \bibnamefont {Zeng}}, \bibinfo {author} {\bibfnamefont {Guofu}\ \bibnamefont
  {Zhou}}, \bibinfo {author} {\bibfnamefont {Jun-Ming}\ \bibnamefont {Liu}},
  \bibinfo {author} {\bibfnamefont {Krzysztof}\ \bibnamefont {Kempa}}, \ and\
  \bibinfo {author} {\bibfnamefont {Jinwei}\ \bibnamefont {Gao}},\ }\bibfield
  {title} {\enquote {\bibinfo {title} {Colossal figure of merit in
  transparent-conducting metallic ribbon networks},}\ }\href {\doibase
  10.1002/admt.201600095} {\bibfield  {journal} {\bibinfo  {journal} {Adv.
  Mater. Technol.}\ }\textbf {\bibinfo {volume} {1}},\ \bibinfo {pages}
  {1600095} (\bibinfo {year} {2016})}\BibitemShut {NoStop}%
\bibitem [{\citenamefont {Voronin}\ \emph {et~al.}(2016)\citenamefont
  {Voronin}, \citenamefont {Ivanchenko}, \citenamefont {Simunin}, \citenamefont
  {Shiverskiy}, \citenamefont {Aleksandrovsky}, \citenamefont {Nemtsev},
  \citenamefont {Fadeev}, \citenamefont {Karpova},\ and\ \citenamefont
  {Khartov}}]{Voronin2016}%
  \BibitemOpen
  \bibfield  {author} {\bibinfo {author} {\bibfnamefont {A.~S.}\ \bibnamefont
  {Voronin}}, \bibinfo {author} {\bibfnamefont {F.~S.}\ \bibnamefont
  {Ivanchenko}}, \bibinfo {author} {\bibfnamefont {M.~M.}\ \bibnamefont
  {Simunin}}, \bibinfo {author} {\bibfnamefont {A.~V.}\ \bibnamefont
  {Shiverskiy}}, \bibinfo {author} {\bibnamefont {Aleksandrovsky}}, \bibinfo
  {author} {\bibfnamefont {I.~V.}\ \bibnamefont {Nemtsev}}, \bibinfo {author}
  {\bibfnamefont {Y.~V.}\ \bibnamefont {Fadeev}}, \bibinfo {author}
  {\bibfnamefont {D.~V.}\ \bibnamefont {Karpova}}, \ and\ \bibinfo {author}
  {\bibfnamefont {S.~V.}\ \bibnamefont {Khartov}},\ }\bibfield  {title}
  {\enquote {\bibinfo {title} {High performance hybrid {rGO}/{Ag}
  quasi-periodic mesh transparent electrodes for flexible electrochromic
  devices},}\ }\href {\doibase 10.1016/j.apsusc.2015.12.182} {\bibfield
  {journal} {\bibinfo  {journal} {Appl. Surf. Sci.}\ }\textbf {\bibinfo
  {volume} {364}},\ \bibinfo {pages} {931--937} (\bibinfo {year}
  {2016})}\BibitemShut {NoStop}%
\bibitem [{\citenamefont {Voronin}\ \emph {et~al.}(2019)\citenamefont
  {Voronin}, \citenamefont {Simunin}, \citenamefont {Fadeev}, \citenamefont
  {Ivanchenko}, \citenamefont {Karpova}, \citenamefont {Tambasov},\ and\
  \citenamefont {Khartov}}]{Voronin2019}%
  \BibitemOpen
  \bibfield  {author} {\bibinfo {author} {\bibfnamefont {A.~S.}\ \bibnamefont
  {Voronin}}, \bibinfo {author} {\bibfnamefont {M.~M.}\ \bibnamefont
  {Simunin}}, \bibinfo {author} {\bibfnamefont {Yu.~V.}\ \bibnamefont
  {Fadeev}}, \bibinfo {author} {\bibfnamefont {F.~S.}\ \bibnamefont
  {Ivanchenko}}, \bibinfo {author} {\bibfnamefont {D.~V.}\ \bibnamefont
  {Karpova}}, \bibinfo {author} {\bibfnamefont {I.~A.}\ \bibnamefont
  {Tambasov}}, \ and\ \bibinfo {author} {\bibfnamefont {S.~V.}\ \bibnamefont
  {Khartov}},\ }\bibfield  {title} {\enquote {\bibinfo {title} {Technological
  basis of the formation of micromesh transparent electrodes by means of a
  self-organized template and the study of their properties},}\ }\href
  {\doibase 10.1134/s1063785019040187} {\bibfield  {journal} {\bibinfo
  {journal} {Tech. Phys. Lett.}\ }\textbf {\bibinfo {volume} {45}},\ \bibinfo
  {pages} {366--369} (\bibinfo {year} {2019})}\BibitemShut {NoStop}%
\bibitem [{\citenamefont {Voronin}\ \emph
  {et~al.}(2021{\natexlab{a}})\citenamefont {Voronin}, \citenamefont {Fadeev},
  \citenamefont {Govorun}, \citenamefont {Voloshin}, \citenamefont {Tambasov},
  \citenamefont {Simunin},\ and\ \citenamefont {Khartov}}]{Voronin2021a}%
  \BibitemOpen
  \bibfield  {author} {\bibinfo {author} {\bibfnamefont {A.~S.}\ \bibnamefont
  {Voronin}}, \bibinfo {author} {\bibfnamefont {Yu.~V.}\ \bibnamefont
  {Fadeev}}, \bibinfo {author} {\bibfnamefont {I.~V.}\ \bibnamefont {Govorun}},
  \bibinfo {author} {\bibfnamefont {A.~S.}\ \bibnamefont {Voloshin}}, \bibinfo
  {author} {\bibfnamefont {I.~A.}\ \bibnamefont {Tambasov}}, \bibinfo {author}
  {\bibfnamefont {M.~M.}\ \bibnamefont {Simunin}}, \ and\ \bibinfo {author}
  {\bibfnamefont {S.~V.}\ \bibnamefont {Khartov}},\ }\bibfield  {title}
  {\enquote {\bibinfo {title} {A transparent radio frequency shielding coating
  obtained using a self-organized template},}\ }\href {\doibase
  10.1134/s1063785021030159} {\bibfield  {journal} {\bibinfo  {journal} {Tech.
  Phys. Lett.}\ }\textbf {\bibinfo {volume} {47}},\ \bibinfo {pages} {259--262}
  (\bibinfo {year} {2021}{\natexlab{a}})}\BibitemShut {NoStop}%
\bibitem [{\citenamefont {Voronin}\ \emph
  {et~al.}(2021{\natexlab{b}})\citenamefont {Voronin}, \citenamefont {Fadeev},
  \citenamefont {Govorun}, \citenamefont {Podshivalov}, \citenamefont
  {Simunin}, \citenamefont {Tambasov}, \citenamefont {Karpova}, \citenamefont
  {Smolyarova}, \citenamefont {Lukyanenko}, \citenamefont {Karacharov},
  \citenamefont {Nemtsev},\ and\ \citenamefont {Khartov}}]{Voronin2021}%
  \BibitemOpen
  \bibfield  {author} {\bibinfo {author} {\bibfnamefont {A.~S.}\ \bibnamefont
  {Voronin}}, \bibinfo {author} {\bibfnamefont {Y.~V.}\ \bibnamefont {Fadeev}},
  \bibinfo {author} {\bibfnamefont {I.~V.}\ \bibnamefont {Govorun}}, \bibinfo
  {author} {\bibfnamefont {I.~V.}\ \bibnamefont {Podshivalov}}, \bibinfo
  {author} {\bibfnamefont {M.~M.}\ \bibnamefont {Simunin}}, \bibinfo {author}
  {\bibfnamefont {I.~A.}\ \bibnamefont {Tambasov}}, \bibinfo {author}
  {\bibfnamefont {D.~V.}\ \bibnamefont {Karpova}}, \bibinfo {author}
  {\bibfnamefont {T.~E.}\ \bibnamefont {Smolyarova}}, \bibinfo {author}
  {\bibfnamefont {A.~V.}\ \bibnamefont {Lukyanenko}}, \bibinfo {author}
  {\bibfnamefont {A.~A.}\ \bibnamefont {Karacharov}}, \bibinfo {author}
  {\bibfnamefont {I.~V.}\ \bibnamefont {Nemtsev}}, \ and\ \bibinfo {author}
  {\bibfnamefont {S.~V.}\ \bibnamefont {Khartov}},\ }\bibfield  {title}
  {\enquote {\bibinfo {title} {Cu--{Ag} and {Ni}--{Ag} meshes based on cracked
  template as efficient transparent electromagnetic shielding coating with
  excellent mechanical performance},}\ }\href {\doibase
  10.1007/s10853-021-06206-4} {\bibfield  {journal} {\bibinfo  {journal} {J.
  Mater. Sci.}\ }\textbf {\bibinfo {volume} {56}},\ \bibinfo {pages}
  {14741--14762} (\bibinfo {year} {2021}{\natexlab{b}})}\BibitemShut {NoStop}%
\bibitem [{\citenamefont {Voronin}\ \emph {et~al.}(2023)\citenamefont
  {Voronin}, \citenamefont {Fadeev}, \citenamefont {Ivanchenko}, \citenamefont
  {Dobrosmyslov}, \citenamefont {Makeev}, \citenamefont {Mikhalev},
  \citenamefont {Osipkov}, \citenamefont {Damaratsky}, \citenamefont
  {Ryzhenko}, \citenamefont {Yurkov}, \citenamefont {Simunin}, \citenamefont
  {Volochaev}, \citenamefont {Tambasov}, \citenamefont {Nedelin}, \citenamefont
  {Zolotovsky}, \citenamefont {Bainov},\ and\ \citenamefont
  {Khartov}}]{Voronin2023}%
  \BibitemOpen
  \bibfield  {author} {\bibinfo {author} {\bibfnamefont {A.~S.}\ \bibnamefont
  {Voronin}}, \bibinfo {author} {\bibfnamefont {Y.~V.}\ \bibnamefont {Fadeev}},
  \bibinfo {author} {\bibfnamefont {F.~S.}\ \bibnamefont {Ivanchenko}},
  \bibinfo {author} {\bibfnamefont {S.~S.}\ \bibnamefont {Dobrosmyslov}},
  \bibinfo {author} {\bibfnamefont {M.~O.}\ \bibnamefont {Makeev}}, \bibinfo
  {author} {\bibfnamefont {P.~A.}\ \bibnamefont {Mikhalev}}, \bibinfo {author}
  {\bibfnamefont {A.~S.}\ \bibnamefont {Osipkov}}, \bibinfo {author}
  {\bibfnamefont {I.~A.}\ \bibnamefont {Damaratsky}}, \bibinfo {author}
  {\bibfnamefont {D.~S.}\ \bibnamefont {Ryzhenko}}, \bibinfo {author}
  {\bibfnamefont {G.~Y.}\ \bibnamefont {Yurkov}}, \bibinfo {author}
  {\bibfnamefont {M.~M.}\ \bibnamefont {Simunin}}, \bibinfo {author}
  {\bibfnamefont {M.~N.}\ \bibnamefont {Volochaev}}, \bibinfo {author}
  {\bibfnamefont {I.~A.}\ \bibnamefont {Tambasov}}, \bibinfo {author}
  {\bibfnamefont {S.~V.}\ \bibnamefont {Nedelin}}, \bibinfo {author}
  {\bibfnamefont {N.~A.}\ \bibnamefont {Zolotovsky}}, \bibinfo {author}
  {\bibfnamefont {D.~D.}\ \bibnamefont {Bainov}}, \ and\ \bibinfo {author}
  {\bibfnamefont {S.~V.}\ \bibnamefont {Khartov}},\ }\bibfield  {title}
  {\enquote {\bibinfo {title} {Original concept of cracked template with
  controlled peeling of the cells perimeter for high performance transparent
  {EMI} shielding films},}\ }\href {\doibase 10.1016/j.surfin.2023.102793}
  {\bibfield  {journal} {\bibinfo  {journal} {Surf. Interfaces}\ }\textbf
  {\bibinfo {volume} {38}},\ \bibinfo {pages} {102793} (\bibinfo {year}
  {2023})}\BibitemShut {NoStop}%
\bibitem [{\citenamefont {Kang}\ \emph {et~al.}(2022)\citenamefont {Kang},
  \citenamefont {Arepalli}, \citenamefont {Yang}, \citenamefont {Lee},
  \citenamefont {Wi}, \citenamefont {Yun}, \citenamefont {Song}, \citenamefont
  {Kim}, \citenamefont {Eo}, \citenamefont {Cho}, \citenamefont {Gwak},\ and\
  \citenamefont {Chung}}]{Kang2022}%
  \BibitemOpen
  \bibfield  {author} {\bibinfo {author} {\bibfnamefont {Seoin}\ \bibnamefont
  {Kang}}, \bibinfo {author} {\bibfnamefont {Vinaya~Kumar}\ \bibnamefont
  {Arepalli}}, \bibinfo {author} {\bibfnamefont {Eunyeong}\ \bibnamefont
  {Yang}}, \bibinfo {author} {\bibfnamefont {Sangyeob}\ \bibnamefont {Lee}},
  \bibinfo {author} {\bibfnamefont {Jung-Sub}\ \bibnamefont {Wi}}, \bibinfo
  {author} {\bibfnamefont {Jae~Ho}\ \bibnamefont {Yun}}, \bibinfo {author}
  {\bibfnamefont {Soomin}\ \bibnamefont {Song}}, \bibinfo {author}
  {\bibfnamefont {Kihwan}\ \bibnamefont {Kim}}, \bibinfo {author}
  {\bibfnamefont {Young-Joo}\ \bibnamefont {Eo}}, \bibinfo {author}
  {\bibfnamefont {Jun-Sik}\ \bibnamefont {Cho}}, \bibinfo {author}
  {\bibfnamefont {Jihye}\ \bibnamefont {Gwak}}, \ and\ \bibinfo {author}
  {\bibfnamefont {Choong-Heui}\ \bibnamefont {Chung}},\ }\bibfield  {title}
  {\enquote {\bibinfo {title} {High performance and flexible electrodeposited
  silver mesh transparent conducting electrodes based on a self-cracking
  template},}\ }\href {\doibase 10.1007/s13391-022-00358-4} {\bibfield
  {journal} {\bibinfo  {journal} {Electron. Mater. Lett.}\ }\textbf {\bibinfo
  {volume} {18}},\ \bibinfo {pages} {440--446} (\bibinfo {year}
  {2022})}\BibitemShut {NoStop}%
\bibitem [{\citenamefont {Cheuk}\ \emph {et~al.}(2016)\citenamefont {Cheuk},
  \citenamefont {Pei},\ and\ \citenamefont {Chan}}]{Cheuk2016}%
  \BibitemOpen
  \bibfield  {author} {\bibinfo {author} {\bibfnamefont {Kin~Wai}\ \bibnamefont
  {Cheuk}}, \bibinfo {author} {\bibfnamefont {Ke}~\bibnamefont {Pei}}, \ and\
  \bibinfo {author} {\bibfnamefont {Paddy K.~L.}\ \bibnamefont {Chan}},\
  }\bibfield  {title} {\enquote {\bibinfo {title} {Degradation mechanism of a
  junction-free transparent silver network electrode},}\ }\href {\doibase
  10.1039/C6RA15135C} {\bibfield  {journal} {\bibinfo  {journal} {RSC Adv.}\
  }\textbf {\bibinfo {volume} {6}},\ \bibinfo {pages} {73769--73775} (\bibinfo
  {year} {2016})}\BibitemShut {NoStop}%
\bibitem [{\citenamefont {Voronin}\ \emph {et~al.}(2022)\citenamefont
  {Voronin}, \citenamefont {Fadeev}, \citenamefont {Makeev}, \citenamefont
  {Mikhalev}, \citenamefont {Osipkov}, \citenamefont {Provatorov},
  \citenamefont {Ryzhenko}, \citenamefont {Yurkov}, \citenamefont {Simunin},
  \citenamefont {Karpova}, \citenamefont {Lukyanenko}, \citenamefont {Kokh},
  \citenamefont {Bainov}, \citenamefont {Tambasov}, \citenamefont {Nedelin},
  \citenamefont {Zolotovsky},\ and\ \citenamefont {Khartov}}]{Voronin2022}%
  \BibitemOpen
  \bibfield  {author} {\bibinfo {author} {\bibfnamefont {Anton~S.}\
  \bibnamefont {Voronin}}, \bibinfo {author} {\bibfnamefont {Yurii~V.}\
  \bibnamefont {Fadeev}}, \bibinfo {author} {\bibfnamefont {Mstislav~O.}\
  \bibnamefont {Makeev}}, \bibinfo {author} {\bibfnamefont {Pavel~A.}\
  \bibnamefont {Mikhalev}}, \bibinfo {author} {\bibfnamefont {Alexey~S.}\
  \bibnamefont {Osipkov}}, \bibinfo {author} {\bibfnamefont {Alexander~S.}\
  \bibnamefont {Provatorov}}, \bibinfo {author} {\bibfnamefont {Dmitriy~S.}\
  \bibnamefont {Ryzhenko}}, \bibinfo {author} {\bibfnamefont {Gleb~Y.}\
  \bibnamefont {Yurkov}}, \bibinfo {author} {\bibfnamefont {Mikhail~M.}\
  \bibnamefont {Simunin}}, \bibinfo {author} {\bibfnamefont {Darina~V.}\
  \bibnamefont {Karpova}}, \bibinfo {author} {\bibfnamefont {Anna~V.}\
  \bibnamefont {Lukyanenko}}, \bibinfo {author} {\bibfnamefont {Dieter}\
  \bibnamefont {Kokh}}, \bibinfo {author} {\bibfnamefont {Dashi~D.}\
  \bibnamefont {Bainov}}, \bibinfo {author} {\bibfnamefont {Igor~A.}\
  \bibnamefont {Tambasov}}, \bibinfo {author} {\bibfnamefont {Sergey~V.}\
  \bibnamefont {Nedelin}}, \bibinfo {author} {\bibfnamefont {Nikita~A.}\
  \bibnamefont {Zolotovsky}}, \ and\ \bibinfo {author} {\bibfnamefont
  {Stanislav~V.}\ \bibnamefont {Khartov}},\ }\bibfield  {title} {\enquote
  {\bibinfo {title} {Low cost embedded copper mesh based on cracked template
  for highly durability transparent {EMI} shielding films},}\ }\href {\doibase
  10.3390/ma15041449} {\bibfield  {journal} {\bibinfo  {journal} {Materials}\
  }\textbf {\bibinfo {volume} {15}},\ \bibinfo {pages} {1449} (\bibinfo {year}
  {2022})}\BibitemShut {NoStop}%
\bibitem [{\citenamefont {Walia}\ \emph {et~al.}(2020)\citenamefont {Walia},
  \citenamefont {Singh}, \citenamefont {Rao}, \citenamefont {Bose},\ and\
  \citenamefont {Kulkarni}}]{Walia2020}%
  \BibitemOpen
  \bibfield  {author} {\bibinfo {author} {\bibfnamefont {Sunil}\ \bibnamefont
  {Walia}}, \bibinfo {author} {\bibfnamefont {Ashutosh~K.}\ \bibnamefont
  {Singh}}, \bibinfo {author} {\bibfnamefont {Veena S.~G.}\ \bibnamefont
  {Rao}}, \bibinfo {author} {\bibfnamefont {Suryasarathi}\ \bibnamefont
  {Bose}}, \ and\ \bibinfo {author} {\bibfnamefont {Giridhar~U.}\ \bibnamefont
  {Kulkarni}},\ }\bibfield  {title} {\enquote {\bibinfo {title} {Metal
  mesh-based transparent electrodes as high-performance {EMI} shields},}\
  }\href {\doibase 10.1007/s12034-020-02159-7} {\bibfield  {journal} {\bibinfo
  {journal} {Bull. Mater. Sci.}\ }\textbf {\bibinfo {volume} {43}},\ \bibinfo
  {pages} {187} (\bibinfo {year} {2020})}\BibitemShut {NoStop}%
\bibitem [{\citenamefont {Liu}\ \emph {et~al.}(2022)\citenamefont {Liu},
  \citenamefont {Huang}, \citenamefont {Peng}, \citenamefont {Liu},
  \citenamefont {Gao},\ and\ \citenamefont {Wang}}]{Liu2022}%
  \BibitemOpen
  \bibfield  {author} {\bibinfo {author} {\bibfnamefont {Ping}\ \bibnamefont
  {Liu}}, \bibinfo {author} {\bibfnamefont {Bing}\ \bibnamefont {Huang}},
  \bibinfo {author} {\bibfnamefont {Lei}\ \bibnamefont {Peng}}, \bibinfo
  {author} {\bibfnamefont {Liming}\ \bibnamefont {Liu}}, \bibinfo {author}
  {\bibfnamefont {Qingguo}\ \bibnamefont {Gao}}, \ and\ \bibinfo {author}
  {\bibfnamefont {Yuehui}\ \bibnamefont {Wang}},\ }\bibfield  {title} {\enquote
  {\bibinfo {title} {A crack templated copper network film as a transparent
  conductive film and its application in organic light-emitting diode},}\
  }\href {\doibase 10.1038/s41598-022-24672-x} {\bibfield  {journal} {\bibinfo
  {journal} {Sci. Rep.}\ }\textbf {\bibinfo {volume} {12}},\ \bibinfo {pages}
  {20494} (\bibinfo {year} {2022})}\BibitemShut {NoStop}%
\bibitem [{\citenamefont {Mondal}\ \emph {et~al.}(2020)\citenamefont {Mondal},
  \citenamefont {Bahuguna}, \citenamefont {Ganesha}, \citenamefont {Verma},
  \citenamefont {Gupta}, \citenamefont {Singh},\ and\ \citenamefont
  {Kulkarni}}]{Mondal2020}%
  \BibitemOpen
  \bibfield  {author} {\bibinfo {author} {\bibfnamefont {Indrajit}\
  \bibnamefont {Mondal}}, \bibinfo {author} {\bibfnamefont {Gaurav}\
  \bibnamefont {Bahuguna}}, \bibinfo {author} {\bibfnamefont {Mukhesh~K.}\
  \bibnamefont {Ganesha}}, \bibinfo {author} {\bibfnamefont {Mohit}\
  \bibnamefont {Verma}}, \bibinfo {author} {\bibfnamefont {Ritu}\ \bibnamefont
  {Gupta}}, \bibinfo {author} {\bibfnamefont {Ashutosh~K.}\ \bibnamefont
  {Singh}}, \ and\ \bibinfo {author} {\bibfnamefont {Giridhar~U.}\ \bibnamefont
  {Kulkarni}},\ }\bibfield  {title} {\enquote {\bibinfo {title} {Scalable
  fabrication of scratch-proof transparent {Al}/{F}--{SnO}$_2$ hybrid
  electrodes with unusual thermal and environmental stability},}\ }\href
  {\doibase 10.1021/acsami.0c17018} {\bibfield  {journal} {\bibinfo  {journal}
  {ACS Appl. Mater. Interfaces}\ }\textbf {\bibinfo {volume} {12}},\ \bibinfo
  {pages} {54203--54211} (\bibinfo {year} {2020})}\BibitemShut {NoStop}%
\bibitem [{\citenamefont {Govind}\ \emph {et~al.}(2022)\citenamefont {Govind},
  \citenamefont {Mondal}, \citenamefont {Baishya}, \citenamefont {Ganesha},
  \citenamefont {Walia}, \citenamefont {Singh},\ and\ \citenamefont
  {Kulkarni}}]{Govind2022}%
  \BibitemOpen
  \bibfield  {author} {\bibinfo {author} {\bibfnamefont {Remya~K.}\
  \bibnamefont {Govind}}, \bibinfo {author} {\bibfnamefont {Indrajit}\
  \bibnamefont {Mondal}}, \bibinfo {author} {\bibfnamefont {Kaushik}\
  \bibnamefont {Baishya}}, \bibinfo {author} {\bibfnamefont {Mukhesh~K.}\
  \bibnamefont {Ganesha}}, \bibinfo {author} {\bibfnamefont {Sunil}\
  \bibnamefont {Walia}}, \bibinfo {author} {\bibfnamefont {Ashutosh~K.}\
  \bibnamefont {Singh}}, \ and\ \bibinfo {author} {\bibfnamefont {Giridhar~U.}\
  \bibnamefont {Kulkarni}},\ }\bibfield  {title} {\enquote {\bibinfo {title}
  {Large-area fabrication of high performing, flexible, transparent conducting
  electrodes using screen printing and spray coating techniques},}\ }\href
  {\doibase 10.1002/admt.202101120} {\bibfield  {journal} {\bibinfo  {journal}
  {Adv. Mater. Technol.}\ }\textbf {\bibinfo {volume} {7}},\ \bibinfo {pages}
  {2101120} (\bibinfo {year} {2022})}\BibitemShut {NoStop}%
\bibitem [{\citenamefont {Ghosh}\ \emph {et~al.}(2010)\citenamefont {Ghosh},
  \citenamefont {Chen},\ and\ \citenamefont {Pruneri}}]{Ghosh2010}%
  \BibitemOpen
  \bibfield  {author} {\bibinfo {author} {\bibfnamefont {D.~S.}\ \bibnamefont
  {Ghosh}}, \bibinfo {author} {\bibfnamefont {T.~L.}\ \bibnamefont {Chen}}, \
  and\ \bibinfo {author} {\bibfnamefont {V.}~\bibnamefont {Pruneri}},\
  }\bibfield  {title} {\enquote {\bibinfo {title} {High figure-of-merit
  ultrathin metal transparent electrodes incorporating a conductive grid},}\
  }\href {\doibase 10.1063/1.3299259} {\bibfield  {journal} {\bibinfo
  {journal} {Appl. Phys. Lett.}\ }\textbf {\bibinfo {volume} {96}},\ \bibinfo
  {pages} {041109} (\bibinfo {year} {2010})}\BibitemShut {NoStop}%
\bibitem [{\citenamefont {Schneider}\ \emph {et~al.}(2016)\citenamefont
  {Schneider}, \citenamefont {Rohner}, \citenamefont {Thureja}, \citenamefont
  {Schmid}, \citenamefont {Galliker},\ and\ \citenamefont
  {Poulikakos}}]{Schneider2016}%
  \BibitemOpen
  \bibfield  {author} {\bibinfo {author} {\bibfnamefont {Julian}\ \bibnamefont
  {Schneider}}, \bibinfo {author} {\bibfnamefont {Patrik}\ \bibnamefont
  {Rohner}}, \bibinfo {author} {\bibfnamefont {Deepankur}\ \bibnamefont
  {Thureja}}, \bibinfo {author} {\bibfnamefont {Martin}\ \bibnamefont
  {Schmid}}, \bibinfo {author} {\bibfnamefont {Patrick}\ \bibnamefont
  {Galliker}}, \ and\ \bibinfo {author} {\bibfnamefont {Dimos}\ \bibnamefont
  {Poulikakos}},\ }\bibfield  {title} {\enquote {\bibinfo {title}
  {Electrohydrodynamic {NanoDrip} printing of high aspect ratio metal grid
  transparent electrodes},}\ }\href {\doibase 10.1002/adfm.201503705}
  {\bibfield  {journal} {\bibinfo  {journal} {Adv. Funct. Mater.}\ }\textbf
  {\bibinfo {volume} {26}},\ \bibinfo {pages} {833--840} (\bibinfo {year}
  {2016})}\BibitemShut {NoStop}%
\bibitem [{\citenamefont {Muzzillo}\ \emph {et~al.}(2020)\citenamefont
  {Muzzillo}, \citenamefont {Reese},\ and\ \citenamefont
  {Mansfield}}]{Muzzillo2020b}%
  \BibitemOpen
  \bibfield  {author} {\bibinfo {author} {\bibfnamefont {Christopher~P.}\
  \bibnamefont {Muzzillo}}, \bibinfo {author} {\bibfnamefont {Matthew~O.}\
  \bibnamefont {Reese}}, \ and\ \bibinfo {author} {\bibfnamefont {Lorelle~M.}\
  \bibnamefont {Mansfield}},\ }\bibfield  {title} {\enquote {\bibinfo {title}
  {Fundamentals of using cracked film lithography to pattern transparent
  conductive metal grids for photovoltaics},}\ }\href {\doibase
  10.1021/acs.langmuir.0c00276} {\bibfield  {journal} {\bibinfo  {journal}
  {Langmuir}\ }\textbf {\bibinfo {volume} {36}},\ \bibinfo {pages} {4630--4636}
  (\bibinfo {year} {2020})}\BibitemShut {NoStop}%
\bibitem [{\citenamefont {Kumar}\ and\ \citenamefont
  {Kulkarni}(2016)}]{Kumar2016}%
  \BibitemOpen
  \bibfield  {author} {\bibinfo {author} {\bibfnamefont {Ankush}\ \bibnamefont
  {Kumar}}\ and\ \bibinfo {author} {\bibfnamefont {G.~U.}\ \bibnamefont
  {Kulkarni}},\ }\bibfield  {title} {\enquote {\bibinfo {title} {Evaluating
  conducting network based transparent electrodes from geometrical
  considerations},}\ }\href {\doibase 10.1063/1.4939280} {\bibfield  {journal}
  {\bibinfo  {journal} {J. Appl. Phys.}\ }\textbf {\bibinfo {volume} {119}},\
  \bibinfo {pages} {015102} (\bibinfo {year} {2016})}\BibitemShut {NoStop}%
\bibitem [{\citenamefont {Kumar}\ \emph {et~al.}(2017)\citenamefont {Kumar},
  \citenamefont {Vidhyadhiraja},\ and\ \citenamefont {Kulkarni}}]{Kumar2017a}%
  \BibitemOpen
  \bibfield  {author} {\bibinfo {author} {\bibfnamefont {Ankush}\ \bibnamefont
  {Kumar}}, \bibinfo {author} {\bibfnamefont {N.~S.}\ \bibnamefont
  {Vidhyadhiraja}}, \ and\ \bibinfo {author} {\bibfnamefont {Giridhar~U.}\
  \bibnamefont {Kulkarni}},\ }\bibfield  {title} {\enquote {\bibinfo {title}
  {Current distribution in conducting nanowire networks},}\ }\href {\doibase
  10.1063/1.4985792} {\bibfield  {journal} {\bibinfo  {journal} {J. Appl.
  Phys.}\ }\textbf {\bibinfo {volume} {122}},\ \bibinfo {pages} {045101}
  (\bibinfo {year} {2017})}\BibitemShut {NoStop}%
\bibitem [{\citenamefont {Tarasevich}\ \emph {et~al.}(2019)\citenamefont
  {Tarasevich}, \citenamefont {Vodolazskaya}, \citenamefont {Eserkepov},\ and\
  \citenamefont {Akhunzhanov}}]{Tarasevich2019}%
  \BibitemOpen
  \bibfield  {author} {\bibinfo {author} {\bibfnamefont {Yuri~Yu.}\
  \bibnamefont {Tarasevich}}, \bibinfo {author} {\bibfnamefont {Irina~V.}\
  \bibnamefont {Vodolazskaya}}, \bibinfo {author} {\bibfnamefont {Andrei~V.}\
  \bibnamefont {Eserkepov}}, \ and\ \bibinfo {author} {\bibfnamefont
  {Renat~K.}\ \bibnamefont {Akhunzhanov}},\ }\bibfield  {title} {\enquote
  {\bibinfo {title} {Electrical conductance of two-dimensional composites with
  embedded rodlike fillers: An analytical consideration and comparison of two
  computational approaches},}\ }\href {\doibase 10.1063/1.5092351} {\bibfield
  {journal} {\bibinfo  {journal} {J. Appl. Phys.}\ }\textbf {\bibinfo {volume}
  {125}},\ \bibinfo {pages} {134902} (\bibinfo {year} {2019})}\BibitemShut
  {NoStop}%
\bibitem [{\citenamefont {Frenkel}\ and\ \citenamefont
  {Eppenga}(1985)}]{Frenkel1985}%
  \BibitemOpen
  \bibfield  {author} {\bibinfo {author} {\bibfnamefont {D.}~\bibnamefont
  {Frenkel}}\ and\ \bibinfo {author} {\bibfnamefont {R.}~\bibnamefont
  {Eppenga}},\ }\bibfield  {title} {\enquote {\bibinfo {title} {Evidence for
  algebraic orientational order in a two-dimensional hard-core nematic},}\
  }\href {\doibase 10.1103/PhysRevA.31.1776} {\bibfield  {journal} {\bibinfo
  {journal} {Phys. Rev. A}\ }\textbf {\bibinfo {volume} {31}},\ \bibinfo
  {pages} {1776--1787} (\bibinfo {year} {1985})}\BibitemShut {NoStop}%
\bibitem [{\citenamefont {Le~Roux}\ \emph {et~al.}(2013)\citenamefont
  {Le~Roux}, \citenamefont {Medjedoub}, \citenamefont {Dour},\ and\
  \citenamefont {R\'{e}za\"{\i}-Aria}}]{Roux2013}%
  \BibitemOpen
  \bibfield  {author} {\bibinfo {author} {\bibfnamefont {Sabine}\ \bibnamefont
  {Le~Roux}}, \bibinfo {author} {\bibfnamefont {Farid}\ \bibnamefont
  {Medjedoub}}, \bibinfo {author} {\bibfnamefont {Gilles}\ \bibnamefont
  {Dour}}, \ and\ \bibinfo {author} {\bibfnamefont {Farhad}\ \bibnamefont
  {R\'{e}za\"{\i}-Aria}},\ }\bibfield  {title} {\enquote {\bibinfo {title}
  {Image analysis of microscopic crack patterns applied to thermal fatigue
  heat-checking of high temperature tool steels},}\ }\href {\doibase
  10.1016/j.micron.2012.08.007} {\bibfield  {journal} {\bibinfo  {journal}
  {Micron}\ }\textbf {\bibinfo {volume} {44}},\ \bibinfo {pages} {347--358}
  (\bibinfo {year} {2013})}\BibitemShut {NoStop}%
\bibitem [{\citenamefont {Goehring}\ \emph {et~al.}(2015)\citenamefont
  {Goehring}, \citenamefont {Nakahara}, \citenamefont {Dutta}, \citenamefont
  {Kitsunezaki},\ and\ \citenamefont {Tarafdar}}]{Goehring2015}%
  \BibitemOpen
  \bibfield  {author} {\bibinfo {author} {\bibfnamefont {Lucas}\ \bibnamefont
  {Goehring}}, \bibinfo {author} {\bibfnamefont {Akio}\ \bibnamefont
  {Nakahara}}, \bibinfo {author} {\bibfnamefont {Tapati}\ \bibnamefont
  {Dutta}}, \bibinfo {author} {\bibfnamefont {So}~\bibnamefont {Kitsunezaki}},
  \ and\ \bibinfo {author} {\bibfnamefont {Sujata}\ \bibnamefont {Tarafdar}},\
  }\href {\doibase 10.1002/9783527671922} {\emph {\bibinfo {title} {Desiccation
  Cracks and Their Patterns}}}\ (\bibinfo  {publisher} {Wiley-VCH Verlag
  GmbH},\ \bibinfo {year} {2015})\ p.\ \bibinfo {pages} {368}\BibitemShut
  {NoStop}%
\bibitem [{\citenamefont {Zeng}\ \emph {et~al.}(2020)\citenamefont {Zeng},
  \citenamefont {Wang},\ and\ \citenamefont {Gao}}]{Zeng2020}%
  \BibitemOpen
  \bibfield  {author} {\bibinfo {author} {\bibfnamefont {Zijing}\ \bibnamefont
  {Zeng}}, \bibinfo {author} {\bibfnamefont {Changhong}\ \bibnamefont {Wang}},
  \ and\ \bibinfo {author} {\bibfnamefont {Jinwei}\ \bibnamefont {Gao}},\
  }\bibfield  {title} {\enquote {\bibinfo {title} {Numerical simulation and
  optimization of metallic network for highly efficient transparent conductive
  films},}\ }\href {\doibase 10.1063/1.5141162} {\bibfield  {journal} {\bibinfo
   {journal} {J. Appl. Phys.}\ }\textbf {\bibinfo {volume} {127}},\ \bibinfo
  {pages} {065104} (\bibinfo {year} {2020})}\BibitemShut {NoStop}%
\bibitem [{\citenamefont {Kim}\ and\ \citenamefont {Truskett}(2022)}]{Kim2022}%
  \BibitemOpen
  \bibfield  {author} {\bibinfo {author} {\bibfnamefont {Jaeuk}\ \bibnamefont
  {Kim}}\ and\ \bibinfo {author} {\bibfnamefont {Thomas~M.}\ \bibnamefont
  {Truskett}},\ }\bibfield  {title} {\enquote {\bibinfo {title} {Geometric
  model of crack-templated networks for transparent conductive films},}\ }\href
  {\doibase 10.1063/5.0092946} {\bibfield  {journal} {\bibinfo  {journal}
  {Appl. Phys. Lett.}\ }\textbf {\bibinfo {volume} {120}},\ \bibinfo {pages}
  {211108} (\bibinfo {year} {2022})}\BibitemShut {NoStop}%
\bibitem [{\citenamefont {Esteki}\ \emph {et~al.}(2023)\citenamefont {Esteki},
  \citenamefont {Curic}, \citenamefont {Manning}, \citenamefont {Sheerin},
  \citenamefont {Ferreira}, \citenamefont {Boland},\ and\ \citenamefont
  {Rocha}}]{Esteki2023}%
  \BibitemOpen
  \bibfield  {author} {\bibinfo {author} {\bibfnamefont {K.}~\bibnamefont
  {Esteki}}, \bibinfo {author} {\bibfnamefont {D.}~\bibnamefont {Curic}},
  \bibinfo {author} {\bibfnamefont {H.~G.}\ \bibnamefont {Manning}}, \bibinfo
  {author} {\bibfnamefont {E.}~\bibnamefont {Sheerin}}, \bibinfo {author}
  {\bibfnamefont {M.~S.}\ \bibnamefont {Ferreira}}, \bibinfo {author}
  {\bibfnamefont {J.~J.}\ \bibnamefont {Boland}}, \ and\ \bibinfo {author}
  {\bibfnamefont {C.~G.}\ \bibnamefont {Rocha}},\ }\bibfield  {title} {\enquote
  {\bibinfo {title} {Thermo-electro-optical properties of seamless metallic
  nanowire networks for transparent conductor applications},}\ }\href {\doibase
  10.1039/d3nr01130e} {\bibfield  {journal} {\bibinfo  {journal} {Nanoscale}\
  }\textbf {\bibinfo {volume} {15}},\ \bibinfo {pages} {10394--10411} (\bibinfo
  {year} {2023})}\BibitemShut {NoStop}%
\bibitem [{\citenamefont {Roy}\ \emph {et~al.}(2022)\citenamefont {Roy},
  \citenamefont {Haque}, \citenamefont {Mitra}, \citenamefont {Tarafdar},\ and\
  \citenamefont {Dutta}}]{Roy2022}%
  \BibitemOpen
  \bibfield  {author} {\bibinfo {author} {\bibfnamefont {A.}~\bibnamefont
  {Roy}}, \bibinfo {author} {\bibfnamefont {R.~A.~I.}\ \bibnamefont {Haque}},
  \bibinfo {author} {\bibfnamefont {A.~J.}\ \bibnamefont {Mitra}}, \bibinfo
  {author} {\bibfnamefont {S.}~\bibnamefont {Tarafdar}}, \ and\ \bibinfo
  {author} {\bibfnamefont {T.}~\bibnamefont {Dutta}},\ }\bibfield  {title}
  {\enquote {\bibinfo {title} {Combinatorial topology and geometry of fracture
  networks},}\ }\href {\doibase 10.1103/physreve.105.034801} {\bibfield
  {journal} {\bibinfo  {journal} {Phys. Rev. E}\ }\textbf {\bibinfo {volume}
  {105}},\ \bibinfo {pages} {034801} (\bibinfo {year} {2022})}\BibitemShut
  {NoStop}%
\bibitem [{Note1()}]{Note1}%
  \BibitemOpen
  \bibinfo {note} {The unpublished photos were kindly provided by A.S.~Voronin
  who is one the authors of that article\cite {Voronin2021}}\BibitemShut
  {NoStop}%
\bibitem [{\citenamefont {Xian}\ \emph {et~al.}(2017)\citenamefont {Xian},
  \citenamefont {Han}, \citenamefont {Li}, \citenamefont {Yang}, \citenamefont
  {Wu}, \citenamefont {Lu}, \citenamefont {Gao}, \citenamefont {Zeng},
  \citenamefont {Wang}, \citenamefont {Bai}, \citenamefont {Naughton},
  \citenamefont {Zhou}, \citenamefont {Liu}, \citenamefont {Kempa},\ and\
  \citenamefont {Gao}}]{Xian2017}%
  \BibitemOpen
  \bibfield  {author} {\bibinfo {author} {\bibfnamefont {Zhike}\ \bibnamefont
  {Xian}}, \bibinfo {author} {\bibfnamefont {Bing}\ \bibnamefont {Han}},
  \bibinfo {author} {\bibfnamefont {Songru}\ \bibnamefont {Li}}, \bibinfo
  {author} {\bibfnamefont {Chaobin}\ \bibnamefont {Yang}}, \bibinfo {author}
  {\bibfnamefont {Sujuan}\ \bibnamefont {Wu}}, \bibinfo {author} {\bibfnamefont
  {Xubing}\ \bibnamefont {Lu}}, \bibinfo {author} {\bibfnamefont {Xingsen}\
  \bibnamefont {Gao}}, \bibinfo {author} {\bibfnamefont {Min}\ \bibnamefont
  {Zeng}}, \bibinfo {author} {\bibfnamefont {Qianming}\ \bibnamefont {Wang}},
  \bibinfo {author} {\bibfnamefont {Pengfei}\ \bibnamefont {Bai}}, \bibinfo
  {author} {\bibfnamefont {Michael~J.}\ \bibnamefont {Naughton}}, \bibinfo
  {author} {\bibfnamefont {Guofu}\ \bibnamefont {Zhou}}, \bibinfo {author}
  {\bibfnamefont {Jun-Ming}\ \bibnamefont {Liu}}, \bibinfo {author}
  {\bibfnamefont {Krzysztof}\ \bibnamefont {Kempa}}, \ and\ \bibinfo {author}
  {\bibfnamefont {Jinwei}\ \bibnamefont {Gao}},\ }\bibfield  {title} {\enquote
  {\bibinfo {title} {A practical {ITO} replacement strategy: Sputtering-free
  processing of a metallic nanonetwork},}\ }\href {\doibase
  10.1002/admt.201700061} {\bibfield  {journal} {\bibinfo  {journal} {Adv.
  Mater. Technol.}\ }\textbf {\bibinfo {volume} {2}},\ \bibinfo {pages}
  {1700061} (\bibinfo {year} {2017})}\BibitemShut {NoStop}%
\bibitem [{\citenamefont {Gao}\ \emph {et~al.}(2018)\citenamefont {Gao},
  \citenamefont {Xian}, \citenamefont {Zhou}, \citenamefont {Liu},\ and\
  \citenamefont {Kempa}}]{Gao2018}%
  \BibitemOpen
  \bibfield  {author} {\bibinfo {author} {\bibfnamefont {Jinwei}\ \bibnamefont
  {Gao}}, \bibinfo {author} {\bibfnamefont {Zhike}\ \bibnamefont {Xian}},
  \bibinfo {author} {\bibfnamefont {Guofu}\ \bibnamefont {Zhou}}, \bibinfo
  {author} {\bibfnamefont {Jun-Ming}\ \bibnamefont {Liu}}, \ and\ \bibinfo
  {author} {\bibfnamefont {Krzysztof}\ \bibnamefont {Kempa}},\ }\bibfield
  {title} {\enquote {\bibinfo {title} {Nature-inspired metallic networks for
  transparent electrodes},}\ }\href {\doibase 10.1002/adfm.201705023}
  {\bibfield  {journal} {\bibinfo  {journal} {Adv. Funct. Mater.}\ }\textbf
  {\bibinfo {volume} {28}},\ \bibinfo {pages} {1705023} (\bibinfo {year}
  {2018})}\BibitemShut {NoStop}%
\bibitem [{\citenamefont {Vecchio}\ \emph {et~al.}(2021)\citenamefont
  {Vecchio}, \citenamefont {Mahler}, \citenamefont {Hammig},\ and\
  \citenamefont {Kotov}}]{Vecchio2021}%
  \BibitemOpen
  \bibfield  {author} {\bibinfo {author} {\bibfnamefont {Drew}\ \bibnamefont
  {Vecchio}}, \bibinfo {author} {\bibfnamefont {Samuel}\ \bibnamefont
  {Mahler}}, \bibinfo {author} {\bibfnamefont {Mark~D.}\ \bibnamefont
  {Hammig}}, \ and\ \bibinfo {author} {\bibfnamefont {Nicholas~A.}\
  \bibnamefont {Kotov}},\ }\href {https://github.com/drewvecchio/StructuralGT}
  {\enquote {\bibinfo {title} {{StructuralGT}},}\ }\bibinfo {howpublished}
  {github} (\bibinfo {year} {2021})\BibitemShut {NoStop}%
\bibitem [{\citenamefont {Guennebaud}\ \emph {et~al.}(2010)\citenamefont
  {Guennebaud}, \citenamefont {Jacob} \emph {et~al.}}]{Guennebaud2010}%
  \BibitemOpen
  \bibfield  {author} {\bibinfo {author} {\bibfnamefont {Ga\"{e}l}\
  \bibnamefont {Guennebaud}}, \bibinfo {author} {\bibfnamefont {Beno\^{i}t}\
  \bibnamefont {Jacob}},  \emph {et~al.},\ }\href {https://eigen.tuxfamily.org}
  {\enquote {\bibinfo {title} {Eigen v3},}\ }\bibinfo {howpublished}
  {https://eigen.tuxfamily.org/} (\bibinfo {year} {2010})\BibitemShut {NoStop}%
\bibitem [{\citenamefont {Gray}\ \emph {et~al.}(1976)\citenamefont {Gray},
  \citenamefont {Anderson}, \citenamefont {Devine},\ and\ \citenamefont
  {Kwasnik}}]{Gray1976}%
  \BibitemOpen
  \bibfield  {author} {\bibinfo {author} {\bibfnamefont {N.~H.}\ \bibnamefont
  {Gray}}, \bibinfo {author} {\bibfnamefont {J.~B.}\ \bibnamefont {Anderson}},
  \bibinfo {author} {\bibfnamefont {J.~D.}\ \bibnamefont {Devine}}, \ and\
  \bibinfo {author} {\bibfnamefont {J.~M.}\ \bibnamefont {Kwasnik}},\
  }\bibfield  {title} {\enquote {\bibinfo {title} {Topological properties of
  random crack networks},}\ }\href {\doibase 10.1007/bf01031092} {\bibfield
  {journal} {\bibinfo  {journal} {J. Int. Ass. Math. Geol.}\ }\textbf {\bibinfo
  {volume} {8}},\ \bibinfo {pages} {617--626} (\bibinfo {year}
  {1976})}\BibitemShut {NoStop}%
\bibitem [{\citenamefont {Xie}\ \emph {et~al.}(2018)\citenamefont {Xie},
  \citenamefont {Li}, \citenamefont {Xu}, \citenamefont {Wang}, \citenamefont
  {Liu},\ and\ \citenamefont {Guo}}]{Xie2018}%
  \BibitemOpen
  \bibfield  {author} {\bibinfo {author} {\bibfnamefont {Shuyao}\ \bibnamefont
  {Xie}}, \bibinfo {author} {\bibfnamefont {Teng}\ \bibnamefont {Li}}, \bibinfo
  {author} {\bibfnamefont {Zijie}\ \bibnamefont {Xu}}, \bibinfo {author}
  {\bibfnamefont {Yanan}\ \bibnamefont {Wang}}, \bibinfo {author}
  {\bibfnamefont {Xiangyang}\ \bibnamefont {Liu}}, \ and\ \bibinfo {author}
  {\bibfnamefont {Wenxi}\ \bibnamefont {Guo}},\ }\bibfield  {title} {\enquote
  {\bibinfo {title} {A high-response transparent heater based on a {CuS}
  nanosheet film with superior mechanical flexibility and chemical
  stability},}\ }\href {\doibase 10.1039/c7nr09667d} {\bibfield  {journal}
  {\bibinfo  {journal} {Nanoscale}\ }\textbf {\bibinfo {volume} {10}},\
  \bibinfo {pages} {6531--6538} (\bibinfo {year} {2018})}\BibitemShut {NoStop}%
\bibitem [{\citenamefont {Bohn}\ \emph
  {et~al.}(2005{\natexlab{a}})\citenamefont {Bohn}, \citenamefont {Pauchard},\
  and\ \citenamefont {Couder}}]{Bohn2005}%
  \BibitemOpen
  \bibfield  {author} {\bibinfo {author} {\bibfnamefont {S.}~\bibnamefont
  {Bohn}}, \bibinfo {author} {\bibfnamefont {L.}~\bibnamefont {Pauchard}}, \
  and\ \bibinfo {author} {\bibfnamefont {Y.}~\bibnamefont {Couder}},\
  }\bibfield  {title} {\enquote {\bibinfo {title} {Hierarchical crack pattern
  as formed by successive domain divisions.}}\ }\href {\doibase
  10.1103/PhysRevE.71.046214} {\bibfield  {journal} {\bibinfo  {journal} {Phys.
  Rev. E}\ }\textbf {\bibinfo {volume} {71}},\ \bibinfo {pages} {046214}
  (\bibinfo {year} {2005}{\natexlab{a}})}\BibitemShut {NoStop}%
\bibitem [{\citenamefont {Bohn}\ \emph
  {et~al.}(2005{\natexlab{b}})\citenamefont {Bohn}, \citenamefont
  {Platkiewicz}, \citenamefont {Andreotti}, \citenamefont {Adda-Bedia},\ and\
  \citenamefont {Couder}}]{Bohn2005a}%
  \BibitemOpen
  \bibfield  {author} {\bibinfo {author} {\bibfnamefont {S.}~\bibnamefont
  {Bohn}}, \bibinfo {author} {\bibfnamefont {J.}~\bibnamefont {Platkiewicz}},
  \bibinfo {author} {\bibfnamefont {B.}~\bibnamefont {Andreotti}}, \bibinfo
  {author} {\bibfnamefont {M.}~\bibnamefont {Adda-Bedia}}, \ and\ \bibinfo
  {author} {\bibfnamefont {Y.}~\bibnamefont {Couder}},\ }\bibfield  {title}
  {\enquote {\bibinfo {title} {Hierarchical crack pattern as formed by
  successive domain divisions. {II}. {From} disordered to deterministic
  behavior},}\ }\href {\doibase 10.1103/PhysRevE.71.046215} {\bibfield
  {journal} {\bibinfo  {journal} {Phys. Rev. E}\ }\textbf {\bibinfo {volume}
  {71}},\ \bibinfo {pages} {046215} (\bibinfo {year}
  {2005}{\natexlab{b}})}\BibitemShut {NoStop}%
\bibitem [{\citenamefont {Kumar}\ and\ \citenamefont
  {Kulkarni}(2021)}]{Kumar2021}%
  \BibitemOpen
  \bibfield  {author} {\bibinfo {author} {\bibfnamefont {Ankush}\ \bibnamefont
  {Kumar}}\ and\ \bibinfo {author} {\bibfnamefont {G.~U.}\ \bibnamefont
  {Kulkarni}},\ }\bibfield  {title} {\enquote {\bibinfo {title} {Time evolution
  and spatial hierarchy of crack patterns},}\ }\href {\doibase
  10.1021/acs.langmuir.1c02363} {\bibfield  {journal} {\bibinfo  {journal}
  {Langmuir}\ }\textbf {\bibinfo {volume} {37}},\ \bibinfo {pages}
  {13141--13147} (\bibinfo {year} {2021})}\BibitemShut {NoStop}%
\end{thebibliography}%

\end{document}